\documentclass{article}

\usepackage{latexsym}

\def\tr{{\rm tr}}
\def\ket#1{\mid~\!\!\!{#1}~\!\!\rangle}
\def\bra#1{\langle~\!\!{#1}~\!\!\!\mid}

\def\r{\buildrel \rho_2 \over \geq }
\def\c{\buildrel \rho_2 \over \leq }

\def\co{\buildrel \rho \over \leq }

\def\cC{{\cal C}}
\def\cT{{\cal T}}
\def\cU{{\cal U}}

\author{\large FEDOR HERBUT \thanks{e-mail:
fedorh@infosky.net}\\ Serbian Academy of Sciences
and Arts, Knez Mihajlova 35,\\ 11000 Belgrade,
Serbia and Montenegro}

\title{\bf \large MUTUAL INFORMATION OF BIPARTITE STATES\\
AND QUANTUM DISCORD\\ IN TERMS OF COHERENCE
INFORMATION}

\begin{document}

\maketitle

\rm \small \noindent In relation of observable
and quantum state, the entity $I_C$ from previous
work quantifies simultaneously coherence,
incompatibility and quantumness. In this article
its application to quantum correlations in
bipartite states is studied. It is shown that
Zurek's quantum discord can always be expressed
as excess coherence information (global minus
local). Strong and weak zero-discord cases are
distinguished and investigated in terms of
necessary and sufficient and sufficient
conditions respectively. A unique string of
relevant subsytem observables, each a function of
the next, for "interrogating" the global state
about the state of the opposite subsystem is
derived with detailed entropy and information
gain discussion. The apparent disappearance of
discord in measurement is investigated, and it is
shown that it is actually shifted from between
subsystems $1$ and $2$ to between subsystems $1$
and $(2+3)$, where $3$ is the measuring
instrument. Finally, it is shown that the global
coherence information $I_C(A_2,\rho_{12})$ is
shifted into the global coherence information
$I_C(A_2,\rho_{123}^f)$ in the final state
$\rho_{123}^f$ of the measurement interaction.

\section{Introduction}

\rm The investigation in this article is directed
at the quantum correlations contained in a
general, i. e., pure or mixed, bipartite state.
By "contained" is meant the von Neumann mutual
information of the state. There are numerous
other important approaches in the literature that
are not limited to the mutual information
\cite{Bennett}. These will not be touched upon in
this work.

No need to expand on the importance of this
problem for quantum information theory, quantum
communications, and quantum computers.

We will distinguish the two subsystems by $1$ and
$2$. The former will be called "the distant"
subsystem, and the latter "the nearby" one. We
will distinguish "local" properties of the nearby
subsystem (or of the distant one), and "global"
ones of the bipartite state.

The approach of this article is based on the
concept of {\it coherence information}. Coherence
of an observable $A$ with respect to a quantum
state $\rho$ and the incompatibility of the two
have been simultaneously quantified by the
concept of {\it coherence information}
$I_C(A,\rho )$ \cite{FHJP05}. It is defined in
three equivalent ways: $$I_C(A,\rho )=
S(\sum_lP_l\rho P_l)-S(\rho ),\eqno{(1a)}$$ where
$\enskip A=\sum_la_lP_l$ is the spectral form of
the Hermitian operator $A$ in terms of distinct
eigenvalues $a_l$, and $S(\dots )$ is the von
Neumann entropy of a state. Further, $$I_C(A,\rho
)=S(\rho ||\sum_lP_l\rho P_l),\eqno{(1b)}$$ where
$S(\rho||\sigma)$ is the relative entropy, a
known function of two states, and finally,
$$S(\rho )= S(A,\rho )+ \sum_lp_lS(P_l\rho
P_l/p_l)- I_C(A,\rho ),\eqno{(1c)}$$ where
$\enskip S(A,\rho )=H(p_l)\equiv
-\sum_lp_llogp_l\enskip$ quantifies the
uncertainty of $A$ in $\rho$ in terms of the
Shannon entropy $H(p_l)$ of the probability
distribution $\enskip \forall l:\enskip p_l\equiv
\tr (P_l\rho ).$

The coherence information $I_C(A,\rho )$
quantifies also the {\it quantumness} in the
relation between observable and state: The
relation is quasi-classical if and only if $A$
and $\rho$ are compatible $[A,\rho]=0$; in this
and only in this case $\enskip I_C(A,\rho )=0.$

It will turn out that the coherence-information
approach of this paper is closely connected with
the Zurek concept of quantum discord. (It will be
called shortly "discord".) Zurek introduced an
approach in which the bipartite state $\rho_{12}$
is investigated by "interrogating" it with a
complete nearby subsystem observable $A_2^c$
\cite{Zurek}, \cite{O-Z}, \cite{ZurekTherm},
\cite{ZurekReview}. The associated discord
$\delta_{A_2^c}(\rho_{12})$ appeared as the
natural quantification of quantumness of the
correlations. It is not entanglement. Also
separable states, which, by definition of
entanglement, do not contain it, are stated to
have positive discord, showing quantumness in
"interrogation" by a concrete subsystem
observable. Nevertheless, discord addresses, just
like entanglement, though in a different way, the
same basic problem of quantum correlations: What
is there typically quantum mechanical in them?

In \cite{ZurekTherm} Zurek takes a
thermodynamical approach to the study of the
physical meaning of least discord $\enskip \check
\delta \equiv inf_{\{A_2^c\}}
\delta_{A_2^c}(\rho_{12}). \enskip$ He does this
using the idea of a quantum demon extracting
locally work from $\rho_{12}$. He finds that
$\check \delta$ equals the (nonnegative) excess
of work that a quantum demon can extract in
comparison with a classical one. He also
discusses how his approach relates to a similar
thermodynamical approach of Oppenheim and the
Horodecki family \cite{O-H}.

In a recent review article the Horodecki family,
Oppenheim et al. gave a detailed presentation on
"local versus non-local information" \cite{H-O-}.
They discuss the connection between their
approach and results with those of Zurek and his
discord. Indirectly, the results of this article
are connected also with this work.

Uhlmann gives an elementary presentation of an
analogous approach to quantum correlations
studies independently from both Zurek and the
Horodecki school of thought \cite{Uhlmann}.

It will be shown that discord is actually
coherence-information excess (global minus
local). This will make it possible to throw new
light on the zero-discord problem. The
"interrogating" complete observable $A_2^c$ will
be generalized to include also incomplete
observables $A_2$. Then a string of relevant
observables, each a function of the next, will be
derived that will eliminate, what will be called,
redundant noise, eliminate the garbled part of
the information gain (on the state of the distant
subsystem), and, finally, eliminate all
quantumness - all this at the cost of diminishing
the information gain.

The state $\enskip \rho_L\equiv \sum_lP_l\rho
P_l\enskip$ appearing in definitions (1a) and
(1b) is the so-called L\"{u}ders mixture of
$\rho$ with respect to $A$ \cite{Lud},
\cite{Auletta} (relation (14.16) on p. 225
there). It is the non-selective (or
entire-ensemble) version. (Some authors call it
"dephasing operation, e. g., \cite{O-H}.) The
admixed L\"{u}ders state is $\enskip P_l\rho
P_l/p_l,\enskip$ where $\enskip p_l\equiv \tr
(\rho P_l)\enskip$ is the corresponding
probability. It appears in the selective (or
definite-result) version (utilized in (1c) e.
g.).

To avoid unnecessary repetitions in the
exposition, the following will be understood
throughout the article: the physical term will be
given priority, like "observable" instead of
"Hermitian operator", "state" instead of
"statistical operator", "mixture" instead of
"decomposition of a statistical operator" (into a
finite or infinite convex combination of
statistical operators), "compatibility" instead
of "commutation", etc.

Observables will be restricted to discrete ones,
and as a rule, given in spectral form like
$\enskip A=\sum_la_lP_l\enskip$ with all
eigenvalues $a_l$ distinct. This will always be
tacitly accompanied by the completeness relation
(decomposition of the identity) $\enskip
\sum_lP_l=1.\enskip$ The sum "$\sum_l$ is finite
or infinite as the case may be. If the sum is
necessarily restricted to be finite for some
claim to be valid, then it will be written, e.
g., like $\enskip \sum_{l=1}^m,\enskip$ and it
will be understood that $m$ is an integer. If the
spectrum may be finite or infinite, we will write
$\enskip \{a_l:\forall l\},\enskip$ etc.; if it
is necessarily finite, we will write $\enskip
\{a_l:l=1,2,\dots ,m\}$. Complete observables
$\enskip A^c=\sum_la_l\ket{l}\bra{l}\enskip$ are
written with the suffix "c".

If the given state $\rho$ has an
infinite-dimensional null space, then also
observables $A$ that have a continuous part in
their spectrum can be considered for
coherence-information studies under the
restriction that the subspace spanned by the
eigen-subspaces of $A$ contains the range of
$\rho$ \cite{Roleof}.

Functions of an observable amount to coarsenings
in the spectrum of the latter. We will prefer the
term "coarsening" because it has the simple
opposite "refinement". Both are order relations
like "smaller or equal" and "larger or equal
respectively".

When an observable $\enskip
A=\sum_la_lP_l\enskip$ and a state $\rho$ are
given, we will speak of "detectable" eigenvalues
$a_l$ or index values $l$ meaning those for which
$\enskip p_l\equiv \tr (\rho P_l)>0$. The
spectrum $\enskip \{a_l:\forall l\},\enskip$ and
the set of the index values $\enskip \{l:\forall
l\}$ will always be understood to be connected by
a fixed one-to-one map, enabling us to talk of
"corresponding" eigenvalue etc.

Mixtures like $\enskip \rho
=\sum_kw_k\rho_k\enskip$, finite or infinite,
will be understood in a formal, not operational
sense, i. e., they express the fact that one can
write $\rho$ in that way. The statistical weights
will be called only "weights"; they can be
positive or zero; in the latter case $\rho_k$
need not be defined, nevertheless by definition
$\enskip w_k\rho_k=0$ (and analogously for other
entities than $\rho_k$). The states $\enskip
\rho_k$ will be referred to as "admixed states".

A mixture $\enskip \rho =\sum_kw_k\rho_k\enskip$
is orthogonal if $\enskip k\not= k'\enskip
\Rightarrow \enskip \rho_k\rho_{k'}=0.\enskip$ An
example is the L\"{u}ders mixture $\enskip
\rho_L= \sum_lp_l(P_l\rho P_l/p_l).\enskip$ Then
the mixing property of entropy is valid: $\enskip
S(\rho_L)=H(p_l)+\sum_lp_lS(P_l\rho
P_l/p_l)\enskip$ (see p. 242 in \cite{Wehrl}).

Both for mixtures and for observables the
subsystem will be exhibited in the index, e. g.,
$\enskip \sum_kw_k\rho_2^k,\enskip
\sum_la_lP_2^l$.

Both in mixtures and in observables we will deal
with coarsenings, and binary relations "linked"
and "chained". To distinguish the two cases, we
will occasionally use the terms "m-coarsening",
"m-linked" and "m-chained" for mixtures, and
"o-coarsening", "o-linked", and "o-chained" for
observables.

Mentioning subsystems, we will often omit
"nearby", and only say "subsystem". One should
note that every general statement is symmetrical
in the sense that one can interchange $1$ and
$2$: the claim is either unchanged or one obtains
the symmetrical equally valid claim. Thus, the
stated choice of nearby and distant is arbitrary.

\section{Role of Coherence
Information in Bipartite Quantum Correlations}

We take a bipartite state $\rho_{12}$ with its
reductions $\enskip \rho_s\equiv
\tr_{s'}\rho_{12},\enskip s,s'=1,2\enskip s\not=
s',\enskip$ and a subsystem observable
$A_2=\sum_la_lP_2^l.\enskip$ Two coherence
informations $\enskip I_C(A_2,\rho_2)\enskip$ and
$\enskip I_C(A_2,\rho_{12})\enskip$ appear. Also
two L\"{u}ders mixtures $\enskip \rho_2^L\equiv
\sum_lP_2^l\rho_2P_2^l\enskip$ and $\enskip
\rho_{12}^L\equiv
\sum_lP_2^l\rho_{12}P_2^l\enskip$ enter the
scene. (Here $P_2^l$ is short for $\enskip
(1\otimes P_2^l).)\enskip$ We utilize the
notation:
$$\forall l:\quad p_l\equiv \tr (\rho_{1
2}P_2 ^l), \eqno{(2a)}$$ $$\forall l,p_l>0:\quad
\rho_{1 2}^l\equiv P_2^l
\rho_{12}P_2^l/p_l,\eqno{(2b)}$$
$$\forall l, p_l>0:\quad \rho_s^l\equiv
\tr_{s'}(\rho_{ 12 }^l),\enskip s,s'=1,2 \enskip
s\not= s'.\eqno{(2c)}$$

Next, we'll need the entropy additivity
accompanying {\it any mixture} $\enskip
\rho=\sum_kw_k\rho_k$:
$$S(\rho )= J+\sum_k \Big(w_kS(\rho_k)\Big),\eqno{(3a)}$$
$$J=\sum_k\Big(w_kS(\rho_k||\rho
)\Big).\eqno{(3b)}$$ (If proof is wanted for the
known relation (3b), see proof of Lemma 4 in
\cite{Mutual}.) If the mixture is orthogonal,
then $J$ takes the special form of the Shannon
entropy $H(w_k)$ due to the mixing property. (See
proposition 7 below for more on $J$.)

Now we consider a relevant decomposition of the
{\it mutual information} $I_{12}\equiv I(\rho_{
12})\equiv S_1+S_2-S_{12}$, where $S_s\equiv
S(\rho_s),\enskip s=1,2,12$.

{\bf Theorem 1:  A)} The mutual information
$I_{12}$ of {\it any} bipartite state $\rho_{
12}$, when viewed in relation to {\it any} given
discrete second-subsystem observable $A_2$, can
be decomposed as follows:

$$I_{12}=
J_{A_2}+\Big(I_C(A_2,\rho_{12})
-I_C(A_2,\rho_2)\Big)+\sum_lp_l
I(\rho_{12}^l),\eqno{(4a)}$$ where
$$J_{A_2}\equiv \sum_lp_lS(\rho_1^l||\rho_1),
\eqno{(4b)}$$ and
$$\rho_1=\sum_lp_l\rho_1^l\eqno{(4c)}$$ is the
{\it distant mixture} induced by $A_2$.

{\bf B)} Each of the three terms on the RHS of
(4a) is {\it always nonnegative}.

{\bf Proof: A)} We utilize the entropy
decompositions (3a) for $\rho_1$ and (1c) for
$\enskip \rho_s,\enskip s=2,12$:
$$I_{12}\equiv S_1+S_2-S_{12}=\Big[J_{A_2}+
\sum_lp_lS(\rho_1^l)\Big]+\Big[H(p_l)+\sum_lp_l
S(\rho_2^l)-I_C(A_2,\rho_2)\Big]-$$
$$\Big[H(p_l)+\sum_lp_lS(\rho_{12}^l)-
I_C(A_2,\rho_{12})\Big]=RHS(4a).$$ This completes
the proof of part A).

{\bf B)} The first and the third terms on the RHS
of (4) are obviously nonnegative. To prove that
also the second term is nonnegative we need two
auxiliary claims.

{\bf Corollary 1:} Decomposition (4a) in
application to the L\"{u}ders mixture
$\rho^L_{12}\equiv \sum_lP_2^l\rho_{12}P_2^l$
gives:
$$I(\rho_{1 2}^L)=
J_{A_2}+\sum_lp_l I(\rho_{12}^l).\eqno{(5)}$$

{\bf Proof:} Straightforward evaluation gives
$\enskip
I_C(A_2,\rho_{12}^L)=I_C(A_2,\rho_2^L)\enskip$
(or see proposition 1 below).\hfill $\Box$

{\bf Lemma 1:} The inequality $\enskip
I(\rho_{12}^L)\leq I(\rho_{12})\enskip$ is always
valid.

{\bf Proof:} As it is well known, the mutual
information can be written in the form of
relative entropy
$I_{12}=S(\rho_{12}||\rho_1\otimes \rho_2)$. By
this same formula also
$I(\rho_{12}^L)=S\Big(\sum_lP_2^l\rho_{12}P_2^l
||\rho_1\otimes (\sum_lP_2^l\rho_2P_2^l)\Big)$.
It was proved by Lindblad \cite{Lind75} for the
finite-dimensional case (Theorem on p. 149 there)
that $S(\Phi \rho ||\Phi \sigma )\leq S(\rho
||\sigma )$ for every two states $\rho$ and
$\sigma$ and every completely positive
trace-preserving map $\Phi$. The inequality was
extended to the infinite-dimensional case by
Uhlmann \cite{Uhlmann2}. (It is unjustly called a
theorem of Uhlmann instead of one of Lindblad and
Uhlmann.)

Since $\Phi \equiv \sum_lP_2^l...P_2^l$ is such a
map, the lemma is proved.\hfill $\Box$\\

{\bf End of proof} of part B) of Theorem 1:
Comparing (4a) and (5) and taking into account
Lemma 1, one obtains
$$I(\rho_{12})-
I(\rho_{12}^L)=I_C(A_2,\rho_{12})-
I_C(A_2,\rho_2)\geq 0$$ in the general
case.\hfill $\Box$\\

In the classical discrete case one has a relation
analogous to (3a) and (3b), and one analogous to
(1c), but the latter with $I_C$ missing. Then a
relation analogous to (4a) is obtained
(analogously as in the proof of theorem 1), but
without the excess of coherence information (the
second term) on the RHS. Following Zurek
\cite{Zurek}, this term {\it quantifies the
quantumness} in the mutual information and is
called the {\it quantum discord} with respect to
a complete or incomplete second-subsystem
discrete observable $A_2$, and it is denoted by
$\delta_{A_2}(\rho_{12})$.

The following {\it physical interpretation} of
(4a) suggests itself. The observable $A_2$ is a
{\it probe} (or an "interrogation", cf
\cite{ZurekTherm}) into the quantum correlations
in $\rho_{12}$ making subsystem $2$ the {\it
nearby} one (the instrument measuring $A_2$
interacts directly with it), and subsystem $1$
the {\it distant} one (no interaction with the
measuring apparatus). Applying (3a) and (3b) to
the mixture (4c), one obtains
$$S(\rho_1)=\sum_l\Big(p_lS(\rho_1^l||\rho_1)\Big)
+\sum_l\Big(p_lS(\rho_1^l)\Big).\eqno{(6)}$$

In view of (6), the first term on the RHS of (4a)
is obviously {\it the information gain} about the
distant subsystem acquired by the probe (cf
\cite{Vedral}, \cite{Zurek}, \cite{O-Z}). The
detectable eigenvalues $a_l$ of $A_2$ distinguish
and enumerate the admixed states $\rho_1^l$, and
the acquired information is the gain in the
distant mixture (4c).

The third term on the RHS of (4a) is the amount
of quantum correlations in $\rho_{12}$ that is
{\it inaccessible} by the probe used. (As easily
seen, it is zero if $A_2$ is complete.) We shall
call it {\it residual correlations}. Both the
first and the third term are entropy terms, i.
e., as easily seen, they are concave with respect
to mixtures. But since the mutual information
appears with a minus sign in the subsystem
entropy decomposition $\enskip
S(\rho_{12})=S(\rho_1)-I(\rho_{12})
+S(\rho_2),\enskip$ the mentioned terms are
actually convex as information quantities should
be.

Discord $\enskip \delta_{A_2}(\rho_{12}),\enskip$
being, in general, {\it excess coherence
information}, i. e., a difference of two
information quantities:
$$\delta_{A_2}(\rho_{12})=
\big[I_C(A_2,\rho_{12})-I_C(A_2,\rho_2)\big],
\eqno{(7)}$$ is neither convex nor concave
(because coherence information is convex, cf
proposition 5 in \cite{FHJP05}). This fact gives
some insight into Lieb's result that mutual
information is neither convex nor concave in the
general case \cite{Lieb}. Some authors call
$I_{12}$ "mutual entropy". Having its behavior
under mixing in view, it is neither information
nor entropy. (See \cite{A-C} - subsection III.c
there - for a different point of view.)

Discord is a necessary accompaniment of the
described probing into $\rho_{12}$ by $A_2$. It
is due to quantumness of the correlations.

Assuming that the observable $A_2=\sum_la_lP_2^l$
is {\it incomplete}, one may wonder how the terms
in (4a) behave when $A_2$ is refined (down to a
complete observable or just to a more complete
one). By {\it refinement} is meant another
observable
$$A_2'=\sum_{l,q}a_{l,q}P_2^{l,q}\eqno{(8a)}$$
(the range of $q$ depends on the value of $l$;
for simplicity, this is omitted in notation). It
is by definition such that it further decomposes
the eigenprojectors of $A$, i. e.,
$$\forall l:\quad P_2^l=\sum
_qP_2^{l,q}.$$

This is refinement in an absolute sense, i. e.,
it does not depend on any state $\rho_2$. We need
the generalization of this notion to {\it
state-dependent refinement}.

Let besides $A_2'$ (cf (8a)) also $A_2$ and
$\rho_2$ be given. Let $l'$ enumerate the
detectable and $l''$ the undetectable eigenvalues
of $A_2$ in $\rho_2$. Then
$$A_2=\sum_{l'}a_{l'}P_2^{l'}+\sum_{l''}a_{l''}
P_2^{l''}.\eqno{(8b)}$$ If $$\forall l':\quad
P_2^{l'}=\sum_qP_2^{l',q},\eqno{(8c)}$$ then we
say that $A_2'$ is a (state-dependent) refinement
of $A_2$ with respect to $\rho_2$, and we write
$A_2'\r A_2$. (The symbol "$\r $" is to remind us
that we are dealing with a reflexive and
transitive binary relation - like "larger or
equal" - that is state dependent.)

{\bf Theorem 2:} In refinement of $A_2$ by $A_2'$
with respect to $\rho_2$ (cf (8a)-(8c)), the
reduction of a given arbitrary bipartite state
$\rho_{12}$, the information gain and the discord
remain equal or become {\it larger}, and the
residual correlations remain the same or become
{\it smaller}. To be explicit quantitatively, one
can write (4a) and (4b) with respect to $A_2'$ as
a two-step expression (as if the probing took
place first with $A_2$, and then it was continued
to $A'_2$):
$$I_{12}=\Bigg \{\sum_l\Big (p_lS(\rho_1^l||\rho_1)\Big
) +\sum_{l,q}\Big
(p_lp_{l,q}S(\rho_1^{l,q}||\rho_1^l)\Big )\Bigg
\}+$$ $$\Bigg \{\delta_{A_2}(\rho_{12})+ \sum_l
p_l\delta_{A_2'}(\rho_{12}^l)\Bigg \}+ \Bigg
\{\sum_{l,q}\Big
(p_lp_{l,q}I(\rho_{12}^{l,q})\Big )
\Bigg\},\eqno{(9)}$$ where the expressions in the
large brackets are the information gain, the
discord and the residual correlations
respectively (and $\enskip p_{l,q}\equiv
\tr(P_2^{l,q}\rho_{12}^l)$).

{\bf Proof} is given in Appendix A.

Information gain is the basic purpose of the
probe, hence, one wants it to be as large as
possible. This is the reason why most studies are
restricted to complete observables $A^c_2$. Then,
whenever $\enskip p_l>0,\enskip$ the state
$\enskip
\ket{l}_2\bra{l}_2\rho_2\ket{l}_2\bra{l}_2/p_l=
\ket{l}_2\bra{l}_2\enskip$ is pure, $\enskip
\rho_{12}^l=\rho_1^l\otimes
\ket{l}_2\bra{l}_2\enskip$ is uncorrelated, and
$\enskip I(\rho_{12}^l)=0.\enskip$ Then (4a) is
simplified to become
$$I(\rho_{12})=J_{A_2^c}+\Big(I_C(A_2^c,\rho_{12})-
I_C(A_2^c,\rho_2)\Big).\eqno{(10)}$$

It was argued in \cite{FHPR02} that taking the
infimum of the discords in (10) (cf (7))
$$\check \delta(\rho_{12})\equiv inf_{\{A_2^c\}}
\delta_{A_2^c} (\rho_{12})\eqno{(11)}$$ one may
obtain {\it an observable-independent quantum
measure of quantumness} in $I_{12}$. Vedral et
al. \cite{Vedral} take into account also
generalized observables, and then, taking the
supremum of the
$J_{A_2^c}=\sum_lp_lS(\rho_1^l||\rho_1)$
expressions, they define the classical part of
$I_{12}$.

\section{On Zero Discord}

As it is obvious from (7), a discord
$\delta_{A_2}(\rho_{12})$ can be {\it zero}
either if both coherence informations are zero,
then we call it {\it strong zero}, or if both
coherence informations are positive but equal. We
call this case {\it weak zero}.

A detailed analysis including open problems (at
least for the author) on unachieved results is
now presented.

\subsection{Strong zero discord with an
incomplete or complete observable}

{\bf Proposition 1:} Each of the following two
relations is a {\it necessary and sufficient
condition} for an observable $\enskip A_2=\sum_l
a_lP_2^l\enskip$ to have a {\it strong zero}
discord in a given bipartite state $\rho_{12}$:
$$[A_2,\rho_{12}]=0,\eqno{(12)}$$
$$\rho_{12}=
\sum_lP_2^l\rho_{12}P_2^l.\eqno{(13)}$$

{\bf Proof:} Upon partial trace over the first
subsystem, the commutation (12) becomes $\enskip
[A_2,\rho_2]=0.\enskip$ Hence the sufficiency and
the necessity of this condition is obvious.

Relation (12) is equivalent to $$\forall l: \quad
[P_2^l,\rho_{12}]=0.\eqno{(14)}$$ The identity
$\enskip
\rho_{12}=(\sum_lP_2^l)\rho_{12},\enskip$
idempotency and commutation then give (13).
Conversely, (13) implies (14).

\hfill $\Box$

{\bf Remark 1:} Relation (12) implies the {\it
local} necessary condition $\enskip
[A_2,\rho_2]=0\enskip$ for strong zero discord. A
local sufficient condition is not possible in a
nontrivial way. Namely, if such a condition were
given in terms of $A_2$ and $\rho_2$ only, one
could make the so-called purification: $\enskip
\rho_{12} \equiv
\ket{\Psi}_{12}\bra{\Psi}_{12}\enskip$ with
$\enskip \tr_1\rho_{12}=\rho_2\enskip$ (the given
local state). Then, relation (14) would imply, as
easily seen,
$$ \exists \bar l:\quad \forall l:\quad (1\otimes
P_2^l) \ket{\Psi}_{12}=\delta_{l,\bar l}
\ket{\Psi}_{12},$$ and further
$$\forall l:\quad P_2^l\rho_2= \delta_{l,\bar l}\rho_2.$$
This gives zero discord, but it also gives zero
information gain $J=0$ because it does not
decompose $\rho_1$
at all, and thus it is a trivial probe.\\

One wants to know what kind of state $\rho_{12}$
has a strong zero discord.

{\bf Definition 1:} If a bipartite state
$\rho_{12}$ is a nontrivial mixture of admixed
states $\rho_{12}^k$
$$\rho_{12}=\sum_kw_k\rho_{12}^k
\eqno{(15a)}$$ (all weights $w_k$ being positive)
so that
$$k\not= k'\quad \Rightarrow \quad
\rho_2^k\rho_2^{k'}=0,\eqno{(15b)}$$ where
$\enskip \forall k:\enskip \rho_2^k\equiv \tr_1
\rho_{12}^k,\enskip$ then $\rho_{12}$ is said to
be {\it mono-orthogonal} (in the second
subsystem).

{\bf Proposition 2:} A bipartite state
$\rho_{12}$ has a strong zero discord {\it if and
only if} it is {\it mono-orthogonal} (in the
second subsystem).

{\bf Proof:} {\it Sufficiency.} Let us assume
that a state $\rho_{12}$ for which (15a) and
(15b) are valid is given. Let us, further, for
each $k$ value denote by $Q_2^k$ the
range-projector of $\rho_2^k$. Finally, let us
define $\enskip A_2 \equiv \sum_ka_kQ_2^k\enskip$
with arbitrary but distinct eigenvalues $a_k$.
Then one has $\enskip \forall k:\enskip
\rho_{12}^k=Q_2^k\rho_{12}^kQ_2^k\enskip$ (This
is a known but not well known general relation.
For proof cf relation (12a) in \cite{FHMD}.)
Hence (14) (changing what has to be changed)
holds true.

{\it Necessity.} If $\rho_{12}$ has a strong zero
discord with respect to an observable $\enskip
A_2=\sum_la_lP_2^l,\enskip$ then, according to
the necessary condition (13), one can write
$\enskip \rho_{12}=\sum_l'p_l
\rho_{12}^l,\enskip$ where the prim on the sum
denotes that all $(p_l=0)$-terms are omitted, and
$\enskip \forall l, p_l>0: \enskip
\rho_{12}^l\equiv P_2^l\rho_{12}P_2^l/p_l$. This
is of the form (15a). Further, $\enskip \forall
l, p_l>0:\enskip \rho_2^l\equiv
\tr_1\rho_{12}^l=P_2^l\rho_2 P_2^l/p_l,$ and
requirement (15b) (with $l$ instead of $k$) is
obviously satisfied. \hfill $\Box$

{\bf Remark 2:} Let it be locally known that
$\rho_{12}$ is mono-orthogonal. This means that
besides $\rho_2$ also an orthogonal projector
decomposition $\enskip \sum_kQ_2^k=Q_2\enskip$ of
the range projector $Q_2$ of $\rho_2$ is given
and it is known that it is associated with
mono-orthogonality, i. e., $Q_2^k$ is the range
projector of $\enskip \rho_2^k\equiv
\tr_1\rho_{12}^k,\enskip$ where $\rho_{12}^k$ are
the admixed mono-orthogonal states in (15a).
Then, as easily seen, a {\it local sufficient
condition} for strong zero discord is that each
eigenprojector $P_2^l$ of $A_2$ be a sum of
$Q_2^k$ projectors. This implies the necessary
condition $\enskip [A_2,\rho_2]=0\enskip$
(because the $Q_2^k$ projectors commute with
$\rho_2$). Nevertheless, it is not a necessary
and sufficient condition, because it may require
too much. A necessary and sufficient local
condition cannot be given in view of lack of
knowledge of the admixed mono-orthogonal states
$\rho_{12}^k$ (cf remark 1).

\subsection{Strong zero discord with a
complete observable}

The necessary and sufficient condition (12) is
unchanged, but, since now $\enskip A_2=\sum_l
a_l\ket{l}_2\bra{l}_2,\enskip$ (13) and (14) take
the respective forms: $$\rho_{12}=\sum_l
\ket{l}_2\bra{l}_2\rho_{12}\ket{l}_2\bra{l}_2,
\eqno{(16)}$$ and $$\forall l:\quad
[\ket{l}_2\bra{l}_2,\rho_{12}]=0.\eqno{(17)}$$

Condition (16) was highlighted in \cite{O-Z} (in
a less elaborate context, without distinguishing
strong and weak zero discord).

{\bf Proposition 3:} A bipartite state
$\rho_{12}$ has a strong zero discord with
respect to a complete observable
$A_2=\sum_la_l\ket{l}_2 \bra{l}_2$ {\it if and
only if} it is a mixture of the form
$$\rho_{12}=\sum_lp_l\rho_1^l\otimes \ket{l}_2
\bra{l}_2.\eqno{(18)}$$

{\bf Proof:} {\it Sufficiency.} If (18) is valid,
then so is (16).

{\it Necessity.} Since $\enskip \forall l:\enskip
\ket{l}_2 \bra{l}_2\rho_{12}\ket{l}_2 \bra{l}_2
=(\bra{l}_2\rho_{12}\ket{l}_2)\ket{l}_2 \bra{l}_2
=p_l\rho_1^l\otimes \ket{l}_2 \bra{l}_2\enskip$
(cf (2b) and (2c) with $\ket{l}_2 \bra{l}_2$
instead of $P_l$). Thus, (16) implies (18).\hfill
$\Box$

{\bf Proposition 4:} A bipartite state
$\rho_{12}$ has a strong zero discord with
respect to some complete observable $A_2$ {\it if
and only if} the state is mono-orthogonal (cf
(15a) and (15b)), and
$$\forall k:\quad \rho_{12}^k=\rho_1^k\otimes
\rho_2^k,$$ i. e., if it is simultaneously also
separable.

{\bf Proof:} {\it Sufficiency.} Let (15a) and
(15b) be given, and let $\rho_{12}$ be
simultaneously also separable as stated.
Substituting each $\rho_2^k$ by its spectral form
in terms of eigen-ray-projectors, one obtains
$\rho_{12}$ as a mixture of the form (18)
(changing what has to be changed).

{\it Necessity.} The form (18) is mono-orthogonal
and simultaneously separable. \hfill $\Box$

{\bf Proposition 5:} Let $\rho_{12}$ be a mixture
of the form
$$\rho_{12}=\sum_kw_k\rho_1^k\otimes \rho_2^k
\eqno{(19)}$$ with the validity of (15b) (cf
definition 1 and proposition 4). Then a {\it
local sufficient condition} for
$A_2=\sum_la_lP_2^l$ to give a strong zero
discord for $\rho_{12}$ is:
$$\forall k:\quad [A_2,\rho_2^k]=0.\eqno{(20)}$$

{\bf Proof:} It is obvious in (19) that, on
account of (20), $A_2$ commutes with $\rho_{12}$
(cf proposition 1). \hfill $\Box$

\subsection{Weak zero discord}

We begin by two general results, which play an
auxiliary role in this subsection.

{\bf Lemma 2:} Let $\rho$ be a state and
$A=\sum_la_lP_l$ an observable. Let, further,
$\enskip \sum_nP_n=1\enskip$ be an (orthogonal
projector) decomposition of the identity such
that $$\forall n:\quad [P_n,\rho
]=[P_n,A]=0.\eqno{(21)}$$ Then the following {\it
statistical decomposition of the coherence
information} ensues: $$I_C(A,\rho )=
\sum_nw_nI_C(A,P_n\rho /w_n),\eqno{(22)}$$ where
$\enskip \forall n:\enskip w_n\equiv \tr (\rho
P_n)$.

{\bf Proof:} On account of (21), one has the
mixture $\enskip \rho =\sum_nw_n(P_n\rho
/w_n),\enskip$ and, $\enskip [P_l,P_n]=0.\enskip$
Hence,
$$I_C(A,\rho )\equiv S\big(\sum_lP_l\rho P_l\big)
-S(\rho)=$$ $$S\big(\sum_nw_n\sum_lP_l(P_n\rho
/w_n)P_l\big)-S\big(\sum_nw_n(P_n\rho
/w_n)\big)=$$ $$H(w_n)+ \sum_nw_nS\big(\sum_lP_l
(P_n\rho /w_n)P_l\big)-$$ $$\big[H(w_n)+\sum_nw_n
S\big(P_n\rho
/w_n\big)\big]=\sum_nw_nI_C(A,P_n\rho /w_n).$$
The symbol $\enskip H(w_n)\enskip$ denotes the
Shannon entropy $\enskip -\tr
(w_nlogw_n),\enskip$ and the mixing property of
entropy has been made use of. \hfill $\Box$

{\bf Lemma 3:} Let $\rho_{12}$ be a bipartite
state and $\enskip A_2=\sum_la_lP_2^l\enskip$ a
subsystem observable. Besides, let $\enskip
\sum_nP_2^n=1\enskip$ be a subsystem (orthogonal
projector) decomposition of the identity such
that $$\forall n:\quad [P_2^n,\rho_{12}]=0\enskip
\mbox{and}\enskip [P_2^n,A_2]=0.\eqno{(23)}$$
Then the following {\it statistical decomposition
of the discord} is valid:
$$\delta_{A_2}(\rho_{12})=
\sum_nw_n\delta_{A_2}(P_2^n\rho_{12} /w_n),\eqno
{(24)}$$ where the mixture $ \rho_{12}=\sum_n[w_n
(P_2^n\rho_{12} /w_n)]$ is due to (23).

{\bf Proof:} Taking the first-subsystem partial
trace in the first commutation relation in (23),
one obtains $\enskip [P_2^n,\rho_2 ]=0.\enskip$
Hence, according to (7) and lemma 2,
$$\delta_{A_2}(\rho_{12})=I_C(A_2,\rho_{12})-
I_C(A_2,\rho_2)=\sum_nw_n\delta_{A_2}(P_2^n
\rho_{12}/w_n).$$\hfill $\Box$

{\bf Proposition 6:} A {\it sufficient condition}
for a weak zero discord of $A_2$ in $\rho_{12}$
is the mixture (19) of the latter with (15b)
valid, further,
$$\forall k:\quad [A_2,Q_2^k], \eqno{(25a)}$$
where $Q_2^k$ is the range projector of
$\rho_2^k$, and finally, for at least one
detectable value $\bar k$ of $k$ one has
$$[A_2,\rho_2^{\bar k}
]\not= 0.\eqno{(25b)}$$

{\bf Proof:} Since $\enskip \forall k:\enskip
\rho_2^k= Q_2^k\rho_2^kQ_2^k,\enskip$ and
$\enskip \rho_1^k\otimes
\rho_2^k=Q_2^k(\rho_1^k\otimes
\rho_2^k)Q_2^k,\enskip$ the assumptions of lemma
3 are satisfied with the decomposition $\enskip
\sum_kQ_2^k=1$. (The null-space projector of
$\rho_2$, if it is nonzero, is joined to the
$Q_2^k$.) Hence, one can write
$$\delta_{A_2}(\rho_{12})=\sum_kw_k\delta_{A_2}
(\rho_1^k\otimes \rho_2^k)=0,$$ because
uncorrelated states have zero mutual information,
and this is an upper bound for the (nonnegative)
discord (cf (7) and (4a)).

On the other hand, also the assumptions of lemma
2 are satisfied. Thus $$I_C(A_2,\rho_2)=\sum_kw_k
I_C(A_2,\rho_2^k)\geq w_{\bar
k}I_C(A_2,\rho_2^{\bar k})>0.$$ In view of (7),
the zero discord must be weak.\hfill $\Box$

{\bf Remark 3:} One would like to know if the
condition in Proposition 6 is also necessary, or
if some other at least partially {\it local}
necessary and sufficient condition is valid.\\

As it is well known, in quantum mechanics, unlike
in the classical discrete case, the von Neumann
mutual information $I_{12}$ can exceed the
subsystem entropies, actually $\enskip I_{12}\leq
2min\big(S(\rho_1),S(\rho_2)\big).\enskip$ Any
correlated pure bipartite state is a good
example, because, as it is also well known, there
$\enskip I_{12}=2S(\rho_1)=2S(\rho_2).\enskip$

Substituting (7) in (4a) and utilizing (3a), (4a)
implies for any complete subsystem observable
$A_2^c$  $$\delta_{A_2^c}=\sum_lp_l
S(\rho_1^l)+(I_{12}-S_1).\eqno{(26)}$$ If
$I_{12}$ {\it exceeds} $S_1$, then (26) gives
rise to a lower bound
$$\delta_{A_2^c}\geq (I_{12}-S_1)>0.\eqno{(27)}$$
Thus, for such typically quantum states
$\rho_{12}$ no choice of $A_2^c$ can give zero
discord.

Cerf and Adami introduced quantum conditional
entropies $S(1|2)$ \cite{C-A}. One has $\enskip
S(1|2)=S_1-I_{12}.\enskip$ If (27) is valid, then
$\enskip S(1|2)<0.\enskip$ It is what Adami and
Cerf call "supercorrelations" \cite{A-C}.

The opposite-sign entity $\enskip -S(1|2)\equiv
E(1\rightarrow 2)\enskip$ is called "directed
entanglement" by Devetak and Staples \cite{D-S}.
Its properties are discussed and it is applied to
quantum communication. The same entity was called
"coherent quantum information" (not to be
confused with "coherence information" of the
present study) by Schumacher and Nielsen
\cite{Sch-N} with analogous discussion and
application.

{\bf Remark 4:} One would like to know if there
can be zero discord between the case of
mono-orthogonal and the case of states for which
(27) is valid. In other words, one wonders if for
some separable but not mono-orthogonal states and
for some nonseparable but states for which
$\enskip I_{12}\leq S_1,\enskip$ one can find a
complete subsystem observable $A_2^c$ giving zero
discord.

{\bf Remark 5:} It is desirable to learn if in
the definition of the least discord $\enskip
\check \delta \equiv
inf_{\{A_2^c\}}\delta_{A_2^c}(\rho_{12})\enskip$
one can replace "inf" by "min" or not. In other
words, it might be that there exist states
$\rho_{12}$ for which $\check \delta$ is
"irrational" in the sense that it can be reached
by no $A_2^c$, but it can be arbitrarily well
approximated by some $\delta_{A_2^c}$. One wants
to see such states if they exist, or to see a
proof that they do not exist. This is, of course,
important also for $\enskip \check \delta =0$.\\

The investigation in this section reveals that
there is a number of open problems
about the zero discord (contrary to a false
impression one might mistakenly get, e. g., from
\cite{O-Z}).

\section{String of Relevant Coarsenings}

\subsection{Elaborate subsystem entropy
decomposition}

When a bipartite state $\rho_{12}$ is given and a
subsystem observable $A_2$ is selected, then the
subsystem entropy decomposition $$S_{12}=S_1-
I_{12}+S_2\eqno{(28a)}$$ can be viewed in the
more elaborate way
$$S_{12}=\Bigg\{\sum_lp_lS(
\rho_1^l)+J_{A_2}(\rho_1)\Bigg\}-\Bigg\{
J_{A_2}(\rho_1)+\delta_{A_2}(\rho_{12})+\sum_lp_l
I(\rho_{12}^l)\Bigg\}+$$ $$
\Bigg\{H(p_l)-I_C(A_2,\rho_2)+\sum_lp_lS(\rho_2^l)
\Bigg\}\eqno{(28b)}$$ (cf (2a)-(2c), (3a) and
(3b), (4a), (7), and (1c)). Naturally, $\enskip
J_{A_2}(\rho_1)=J_{A_2}(\rho_{12}).\enskip$ It is
understood that each expression in large brackets
in (28b) equals the corresponding term on the RHS
of (28a).

{\it The elaborate subsystem entropy
decomposition} (28b) can be {\it interpreted
physically} as follows. The subsystem observable
$A_2$ is chosen to "interrogate" the uncertainty
in the distant subsystem $1$; the measure of the
latter is $S_1$. On account of this, $S_1$ is
broken up into $\enskip \sum_lp_lS(
\rho_1^l),\enskip$ the part of $S_1$ that is
inaccessible to our "interrogation" (or the
residual part), and $\enskip J_{A_2}(\rho_1),
\enskip$ the {\it information gain}. The mutual
information $I_{12}$, which quantifies the total
quantum correlations in $\rho_{12}$, is
decomposed into the mentioned information gain
$\enskip J_{A_2}(\rho_1),\enskip$ the discord
$\enskip \delta_{A_2}(\rho_{12}),\enskip$ and
$\enskip \sum_lp_l I(\rho_{12}^l),\enskip$ which
is the part that is not made use of in the chosen
"interrogation" (residual correlations). The
appearance of the information gain in $I_{12}$
shows that the quantum correlations in
$\rho_{12}$ act as an information channel,
transferring the information gain from subsystem
$1$ to subsystem $2$. The discord appears
because, unless $A_2$ is compatible with
$\rho_{12}$, there is a part of the correlations
that is unsuitable for the mentioned transfer of
the information gain, which is a quasi-classical
notion. This is why it is said that the discord
quantifies the quantumness of the correlations
(regarding $A_2$). Finally, the uncertainty in
$\rho_2$, i. e., $S_2$ is broken up into $\enskip
H(p_l)\equiv
-\sum_lp_llogp_l=S(A_2,\rho_2),\enskip$ the
entropy (or amount of uncertainty) of $A_2$ in
the state of the second subsystem; into the
coherence or incompatibility information $\enskip
I_C(A_2,\rho_2),\enskip$ which is again a
necessary accompaniment of our "interrogation"
due to the quantumness of $\rho_2$; and into
$\enskip \sum_lp_lS(\rho_2^l),\enskip$ which is
the amount of uncertainty in $\rho_2$
inaccessible to $A_2$ (residual uncertainty).

It should be noted that (28b) does not describe a
process; it only gives {\it a relevant
quantitative view} of $\rho_{12}$. In other
words, what the quantum correlations in
$\rho_{12}$ do, among other things, is to
transfer the information gain $J_{A_2}(\rho_1)$
from subsystem $1$ to subsystem $2$. Now it is
natural to ask how we can extract it from
subsystem $2$. Evidently, the thing to do is to
measure $A_2$ on the nearby subsystem $2$, i. e.,
locally (see section VI). But then one extracts
the amount of information $H(p_l)$, and not
$J_{A_2}(\rho_1)$. This motivates the rest of
investigation in this section.

\subsection{Information gain $J$}

It is the aim of this subsection to understand
how the uncertainty $\enskip
H(p_l)=S(A_2,\rho_2)\enskip$ and the information
gain $J_{A_2}(\rho_1)$ relate to each other. We
begin by a precise understanding of $\enskip
J_{A_2}(\rho_1).$

{\bf Proposition 7:} If $\enskip \rho
=\sum_{l=1}^mp_l\rho^l$ is an arbitrary mixture
of a finite number of admixed states, then (3a)
and (3b) are valid. Besides,
$$0\leq J(\rho ) \leq H(p_l),\eqno{(29)}$$ and
$\enskip J(\rho )=0 \enskip$ if and only if
$\enskip \forall l, p_l>0:\enskip
\rho^l=\rho\enskip$ (total overlap), and $\enskip
J(\rho )=H(p_l)\enskip$ if and only if $\enskip
\forall (l\not= l'),\enskip p_l>0<p_{l'}:\enskip
\rho^l\rho^{l'}=0\enskip$ (pairwise
orthogonality).

{\bf Proof:} The first inequality in (29) is
obvious from (3b). The second one is proved in
the review article of Wehrl \cite{Wehrl}
(relation (2.3) there).

One has $\enskip J=0\enskip$ if and only if in
(3b) (changing what has to be changed) $\enskip
p_l>0\enskip \Rightarrow \enskip S(\rho^l||\rho
)=0.\enskip$ It is well known that relative
entropy is zero if and only if the two states in
it coincide.

It is standard knowledge that the so-called
mixing property holds true: if the admixed states
$\rho^l$ are pairwise orthogonal, then $\enskip
J=H(p_l).\enskip$ The converse statement, that
$\enskip J=H(p_l)\enskip$ implies orthogonality
of the $\rho^l$, is not proved anywhere known to
the author of this study. Therefore, its somewhat
lengthy proof, through auxiliary lemmata, is
given in Appendix B.\hfill $\Box$\\

The quantity $\enskip H(p_l)\enskip$ is called
the mixing entropy of the mixture at issue. But
it is only the upper possible extreme value of
the information gain $J$. It is obvious from
proposition 7 that {\it the excess} $\enskip
(H(p_l)-J)\enskip$ (or how much is missing in the
information gain) {\it quantifies the overlap} of
the admixed states. It is zero if and only if
there is no overlap (the admixed states are
orthogonal). It is maximal, i. e., equal to
$\enskip H(p_l),\enskip$ in case of total
overlap, when one is dealing only with an
apparent mixture.

{\bf Remark 6:} It is desirable to have the
extension of proposition 7 to the case of
infinitely many admixed states.\\

To clarify what is apparent and what is genuine
in a mixture, we consider two trivial lemmata.

{\bf Lemma 4:} Let us take a {\it mixture}
$$\rho =\sum_sp_s\rho^s,\quad S(\rho
)=\sum_sp_sS(\rho^s)+J,\eqno{(30a)}$$ and a {\it
refinement} of it
$$\forall s, p_s>0:\quad
\rho^s=\sum_{k_s}w_{k_s}\rho^{k_s},\quad \rho
=\sum_s\sum_{k_s}p_sw_{k_s}\rho^{k_s}.\eqno{(30b)}$$
Then the residual entropy is non-increasing,
whereas the information gain and the mixing
entropy are non-decreasing. More precisely (in
obvious notation):
$$ S(\rho )=\sum_s\sum_{k_s}p_sw_{k_s}S(\rho^{k_s})+
\big\{\sum_s(p_sJ^s)+J\big\},\eqno{(31a)}$$
$$H(p_sw_{k_s})=H(p_s)+\sum_sp_sH(w_{k_s}).\eqno{(31b)}$$

{\bf Proof} is straightforward.

{\bf Lemma 5:} If the refinement in a mixture is
done through mere repetition, i. e., if $\enskip
\forall s,p_s>0:\enskip k_s\not= k'_s\enskip
\Rightarrow \enskip
\rho^{k_s}=\rho^{k'_s},\enskip$ then the residual
entropy and {\it the information gain remain the
same}.

{\bf Proof} is obvious from (31a) if one takes
into account that $\enskip \forall s,
p_s>0:\enskip J^s=0$. \hfill $\Box$

 It is now seen that the information
gain is insensitive to apparent mixing (or
repetition of the admixed states); it depends
only on the genuine mixing, i. e., on the
distinct admixed states. Contrariwise, the mixing
entropy is insensitive to the distinction between
genuine and apparent mixing, i. e., it increases
whenever at least one of the refined probability
distributions is nontrivial. Therefore, in spite
of the fact that $\enskip
\big(H(p_l)-J\big)\enskip$ does quantify the
overlap in the given mixture, which may contain
repetition of admixed states, it can be
diminished on the basis of (31b).

\subsection{Essential noise and garbled
information}

{\bf Definition 2:} If a given mixture $\enskip
\rho =\sum_lp_l\rho_l\enskip$ is rewritten
without repetition of the admixed states with the
use of a new index $s$, the expression $\enskip
\big(H(p_s)-J\big)\enskip$ quantifies the {\it
essential overlap} in the mixture, i. e., the one
due to the genuine mixing of the distinct admixed
states. The original quantity of overlap is the
sum of the quantity of essential overlap and of
that of {\it redundant overlap}: $\enskip
\big(H(p_l)-J\big)= \big(H(p_s)-J\big)+
\big(H(p_l)-H(p_s)\big)$.\\

One can see in (31b) that $\enskip
\big(H(p_l)-H(p_s)\big)\enskip$ is the increase
in the mixing entropy due to repetition of
admixed states.\\

Returning to the elaborate subsystem entropy
decomposition (28b), we see that at best we can
extract the information gain $\enskip
H(p_l)\enskip$ from subsystem $2$ by measuring
the subsystem observable $A_2$ (which is
simultaneously the measurement of $\enskip
(1\otimes A_2)\enskip$ in the bipartite state
$\rho_{12}$). The difference $\enskip \big(H(p_l)
-J_{A_2}\big),\enskip$ corresponding to the
overlap in the distant mixture $\enskip \rho_1
=\sum_lp_l\rho_1^l,\enskip$ appears now as {\it
noise}. In accordance with definition 2, this
noise is the sum of an essential term and a
redundant term. One cannot eliminate the former
(without changing drastically $A_2$, i. e.,
without taking another subsystem observable that
is not a function of $A_2$) because the essential
term is due to the overlap of the distinct
admixed states in $\rho_1$, but one can dispose
of the redundant noise by sheer coarsening.

{\bf Theorem 3:} There exists one and only one
coarsening $B_2^{ess}$ of $A_2$ in which the
redundant noise is and the essential noise is not
eliminated, and the induced distant mixture
$\enskip \rho_1=\sum_sp_s\rho_1^s\enskip$ is
equal to the one obtained due to $A_2$ but
rewritten with positive weights and without
repetitions in the admixed states. To obtain the
subsystem observable $B_2^{ess}$, one defines the
following equivalence relation in the detectable
spectrum of $A_2$: $l\sim l'$ if $\enskip
\rho_1^l=\rho_1^{l'}\enskip$ (cf (2c)). Further,
enumerating by $s$ the obtained equivalence
classes $\enskip \{\cC_s:\forall s\},\enskip$ one
defines
$$B_2^{ess}\equiv \sum_sb_sP_2^s,\eqno{(32a)}$$ where
$\enskip \{b_s:\forall s\}\enskip$ is an
arbitrary set of distinct nonzero real numbers,
and
$$\forall s:\quad P_2^s\equiv \sum_{l\in \cC_s}P_2^l.
\eqno{(32b)}$$

{\bf Proof:} Since $\forall s:\enskip p_s\equiv
\tr (\rho_{12}P_2^s)=\Big(\sum_{l\in
\cC_s}p_l\Big)>0,\enskip$ and $$\rho_1^s\equiv
p_s^{-1}\tr_2 (\rho_{12}P_2^s)=\sum_{l\in
\cC_s}(p_l/p_s)p_l^{-1} \tr_2(\rho_{12}P_2^l)=$$
$$\sum_{l\in \cC_s}(p_l/p_s) \rho_1^l=\rho_1^{\bar l},$$
where $\bar l$ is any index value from the class
$\cC_s$. Thus, $\enskip B_2^{ess}\enskip$ does
induce the desired mixture for $\rho_1$. It is
evidently the unique coarsening of $A_2$  doing
this because every coarsening has to break up the
detectable spectrum of $A_2$ into classes, and
the desired purpose cannot be achieved in any
other way.\hfill $\Box$\\

In general, the information gain $J$ is {\it
garbled} because in the measurement of $A_2$ it
appears necessarily with (inseparable) essential
noise $\enskip (H(p_s)-J)$. (For a precise
definition of "garbled information gain" see the
last but one term in (38) below.)

Needles to say that the expounded procedure of
eliminating redundant noise is analogous in the
classical discrete case of probability
distributions.

\subsection{Orthogonal distant mixture,
pure information gain and twin observables}

As it is obvious from proposition 7 and (28b), if
the distant mixture $\enskip
\rho_1=\sum_{l=1}^mp_l\rho_1^l\enskip$ is {\it
orthogonal}, and only in this case, the essential
noise is zero. Then, one has {\it pure
information}: $\enskip
J_{A_2}=H(p_l)=S(A_2,\rho_2).\enskip$ In this
case there is no redundant noise either. It may
happen that orthogonality is achieved only after
disposing of the redundant noise. Therefore, we
concentrate on $\enskip B_2^{ess}=\sum_sb_sP_2^s
\enskip$ and the corresponding distant mixture
$\enskip \rho_2=\sum_sp_s\rho_2^s\enskip$, but to
make the results more general, the suffix "ess"
is omitted.

Let $Q_1^s$ be the range projector of $\rho_1^s$.
Orthogonality of the above mixture amounts to
$$Q_1^sQ_1^{s'}=\delta_{s,s'}Q_1^s,\eqno{(33a)}$$
and one has $$\sum_sQ_1^s=Q_1,\eqno{(33b)}$$
where $Q_1$ is the range projector of the distant
state $\rho_1$. In this case, we prove the
following result.

{\bf Proposition 8:} Assuming positivity of all
the probabilities $p_s$ and the validity of
$\enskip
\Big(\sum_sP_2^s\Big)\rho_2=\rho_2,\enskip$ if
the {\it distant mixture} $\enskip
\rho_1=\sum_sp_s\rho_1^s\enskip$ (cf (2a)-(2c)
changing what has to be changed) is {\it
orthogonal}, then
$$\Big(\sum_sP_2^s\Big)\rho_{12}=\rho_{12}=
\Big(\sum_sQ_1^s \Big)\rho_{12},\eqno{(34a)}$$
and
$$\forall s:\quad Q_1^s\rho_{12}=P_2^s\rho_{12}
\eqno{(34b)}$$ are valid.

{\bf Proof:} Let $Q_2$ be the range projector of
the nearby state $\rho_2$. The relation $\enskip
(\sum_sP_2^s)\rho_2=\rho_2\enskip$ then implies
$\enskip \Big(\sum_sP_2^s\Big)Q_2=Q_2\enskip$
(see Appendix A in \cite{Roleof}). Since one can
always write $\enskip
\rho_{12}=Q_2\rho_{12}\enskip$ (cf relation (12a)
in \cite{FHMD}), the first equality in (34a)
follows.

The relation (33b) and the fact that one can
write $\enskip \rho_{12}=Q_1\rho_{12}, \enskip$
then make also the second equality in (34a) seen
to be valid.

Next, we prove that $$s\not= s',\quad \Rightarrow
\enskip Q_1^sP_2^{s'}\rho_{12}=0.\eqno{(35)}$$
For unequal index values one has $\enskip \tr
(Q_1^sP_2^{s'}\rho_{12})=p_s\tr (Q_1^s
\rho_1^{s'})=p_s\tr \Big(Q_1^s(Q_1^{s'}
\rho_1^{s'})\Big)=0.\enskip$ Further, $\enskip 0=
\tr (Q_1^sP_2^{s'}\rho_{12})=\tr
\Big((Q_1^sP_2^{s'})\rho_{12}(Q_1^sP_2^{s'})\Big),
\enskip$ and $\enskip
(Q_1^sP_2^{s'})\rho_{12}(Q_1^sP_2^{s'})=0\enskip$
is well known to ensue. Then, the Lemma of
L\"{u}ders (\cite{Lud} or see FN 16 in
\cite{FH69}) entails the claimed relation (35).

Finally, utilizing relations (34a) and (35), one
can argue as follows. $\enskip Q_1^s
\rho_{12}=Q_1^s
\Big(\sum_{s'}P_2^{s'}\Big)\rho_{12}= Q_1^s
P_2^s\rho_{12}=P_2^s\Big(\sum_{s'}Q_1^{s'}\Big)
\rho_{12}=P_2^s\rho_{12}\enskip$ as claimed in
(34b).\hfill $\Box$

If one defines a first-subsystem observable
$\enskip B_1\equiv \sum_sb_sQ_1^s\enskip$ with
arbitrary but distinct nonzero detectable
eigenvalues $\enskip \{b_s:\forall s\},\enskip$
then, according to Theorem 1 in \cite{Roleof} and
the theorem on so-called twin observables (p.
052321-3 in \cite{FHPR02}) imply that proposition
8, actually, gives one more necessary and
sufficient condition for $\enskip
(B_1,B_2)\enskip$ to be {\it twin observables} in
$\rho_{12}$.

Twin observables have a number of remarkable
properties (cf also \cite{FHMD} and the
references therein). For this study an important
implication is that $\enskip
[B_i,\rho_i]=0,\enskip i=1,2\enskip$ (cf the
mentioned Theorem 1 in \cite{Roleof}).

Two obvious consequences on the elaborate
subsystem entropy decomposition (28b), which is
the basic object of this study, follow:

$$I_C(B_2,\rho_2)=0=I_C(B_1,\rho_1),\eqno{(36a)}$$
and, on account of (7), $\enskip
\delta_{A_2}(\rho_{12})=I_C(B_2,\rho_{12}).\enskip$

Thus, in this case (28b) simplifies to
$$S_{12}=S_1-I_{12}+S_2=
\Bigg\{\sum_sp_sS(\rho_1^s)+H(p_s)\Bigg\}-$$ $$
\Bigg\{H(p_s) +I_C(B_2,\rho_{12})+\sum_sp_sI(
\rho_{12}^s)\Bigg\}+\Bigg\{H(p_s)+
\sum_sp_sS(\rho_2^s)\Bigg\},$$ where the mixing
property is utilized for the orthogonal mixture
$\enskip \rho_1=\sum_sp_s\rho_s$.

If $\enskip I_C(B_2,\rho_{12})>0,\enskip$ then we
have the case of so-called {\it correlations
incompatibility} (cf Section 6 in \cite{Roleof}),
in which the discord equals the coherence or
incompatibility information of $B_2$ in
$\rho_{12}$. Besides, there is no quantumness in
$\rho_2$ with respect to $B_2$. (One has global
coherence without local coherence.)

The quantity of uncertainty $S(\rho_2)$ of the
nearby subsystem state now (possibly) exceeds the
quantity of uncertainty $S(B_2,\rho_2)$ of the
obsevable $B_2$ in $\rho_2$, which equals the
pure information gain $\enskip H(p_s)=J_{B_2}.$\\

The assumption $\enskip
\Big(\sum_sP_2^s\Big)\rho_2=\rho_2\enskip$ is
satisfied for $B_2^{ess}$ due to the very
definition of the indices $s$ (all detectable $l$
values of $A_2$ are used up in it). Besides, on
account of the definition of $B_2^{ess}$, all
probabilities $p_s$ are positive.

So far in this subsection we had in mind the
special case when the distant mixture $\enskip
\rho_1=\sum_sp_s\rho_1^s\enskip$ without
repetition in the admixed states turns out
orthogonal. Now we return to the general case and
prove that there always exists a (possibly
trivial) unique minimal coarsening $\enskip
C_2=\sum_tc_tP_2^t\enskip$ of $B_2^{ess}$, and,
by consequence, of $A_2$, that gives an
orthogonal distant mixture and, by a definition
analogous to the above of $B_1$, an observable
$\enskip C_1=\equiv \sum_tc_t'Q_1^t \enskip$ that
is its twin observable.

\subsection{Minimal orthogonal coarsening of a mixture}

Before we proceed, we first expound some relevant
properties of mixtures as far as orthogonal
coarsenings of them are concerned.

{\bf Lemma 6:} For any two states $\rho$ and
$\rho'$ one has $\enskip \tr (\rho\rho')\geq 0,
\enskip$ and $\enskip \tr (\rho\rho')=0\enskip$
if and only if $\enskip \rho\rho'=0$.

{\bf Proof:} Always $\enskip \tr (\rho\rho')=\tr
(\rho^{1/2}\rho' \rho^{1/2})\geq 0\enskip$
because $\enskip \rho^{1/2}\rho'
\rho^{1/2}\enskip$ is a positive operator.
Sufficiency of orthogonality for trace
orthogonality is obvious. Necessity is seen as
follows: $\enskip \tr (\rho\rho')=0\enskip$
implies $\enskip \rho^{1/2}\rho'
\rho^{1/2}=0,\enskip$ and this has, due to the
Lemma of L\"{u}ders (\cite{Lud}), $\enskip
0=\rho'\rho^{1/2} =\rho'\rho \enskip$ as its
consequence.\hfill $\Box$

{\bf Definition 3:} Let $\enskip \rho =\sum_kw_k
\rho_k\enskip$ be a mixture with positive weights
and without repetitions of the admixed states
$\rho_k$. We say that the states $\rho_k$ and
$\rho_{k'}$ are {\it linked} if $\enskip \tr
(\rho_k\rho_{k'})>0.\enskip$ If $\rho_k$ and
$\rho_{k'}$ are such that there exists an integer
$\enskip n,\quad n=1,\enskip \mbox{or}\enskip
2,\enskip \mbox{or}\enskip \dots ,\enskip$ and
there can be found a chain of admixed states
$\enskip \{\rho_{k_i}:i=1,2,\dots ,n\}\enskip$
such that $\enskip \rho_k=\rho_{k_1},\enskip$
$\enskip \rho_{k_n}=\rho_{k'},\enskip$ and any
two neighboring states in the chain are linked,
then we say that $\rho_k$ and $\rho_{k'}$ are
{\it chained}, and we speak of m-chaining.

{\bf Definition 4:} We say that a mixture $\rho
=\sum_tw_t\rho_t \enskip$ is a {\it coarsening}
of another mixture $\enskip \rho
=\sum_sw_s\rho_s,\enskip$ the latter being
without repetition in the admixed states and with
positive weights, if the index set $\enskip
\{s:\forall s\}\enskip$ is partitioned into
m-classes $\cT_t$: $\enskip \{s:\forall
s\}=\sum_t\cT_t\enskip$ (the sum stands for the
union of the non-overlapping classes), is
enumerated by $t$, and $\enskip \rho_t=\sum_{s\in
\cT_t}\bigg(w_s/w_t\bigg)\rho_s,\enskip$ where
$\enskip \forall t:\enskip w_t\equiv
\Big(\sum_{s\in \cT_t}w_s\Big)$. In this case we
speak of m-coarsening.

{\bf Proposition 9:} Let $\enskip \rho =\sum_sp_s
\rho_s\enskip$ be a mixture with all weights
positive and without repetition. Let, further,
another mixture $\enskip \rho
=\sum_tw_t\rho_t\enskip$ be a {\it coarsening} of
the former mixture, obtained by {\it chaining}
(chained m-coarsening). Then the latter mixture
is {\it orthogonal}, and it is {\it minimal} as
such, i. e., if also $\enskip \rho
=\sum_up_u\rho_u$ is an orthogonal coarsening of
the initial mixture, then it is also a coarsening
of its chained m-coarsening.

{\bf Proof:} Orthogonality can be proved as
follows. Let $\enskip t\not= t',\enskip$ and let
$\enskip s\in \cT_t,\enskip$ and $\enskip s'\in
\cT_{t'}.\enskip$ We assume {\it ab contrario}
that $\enskip \tr (\rho_s\rho_{s'})>0.\enskip$
Then, according to definitions 4 and 3, $\rho_s$
and $\rho_{s'}$ are linked, and hence belong to
the same m-class $\cT_t$ contrary to assumption.
Hence, $\enskip \rho^s\rho^{s'}=0 \enskip$ (cf
lemma 6), implying $\enskip \rho^t\rho^{t'}=0
\enskip$ (cf definition 4).

Minimality is proved in the following way. Let
the partitioning $\enskip \{s:\forall
s\}=\sum_u\cU_u\enskip$ define an orthogonal
coarsening $\enskip \rho =\sum_up_u \rho^u
\enskip$ in analogy with definition 4.
Considering the initial mixture $\enskip \rho
=\sum_sp_s\rho^s,\enskip$ we assume that two
distinct index values $s,s'$ are m-linked (cf
definition 3). Lemma 6 claims that $\rho^s$ and
$\rho^{s'}$ then cannot be orthogonal; hence $s$
and $s'$ must belong to one and the same m-class
$\cU_u$. Next, let $s$ and $s'$ be chained. Then
any two neighboring index values in the chain
belong to one and the same m-class $\cU_u$,
entailing the fact that also $s$ and $s'$ belong
to the same m-class. Thus, any m-class $\cT_t$ is
a subset of some m-class $\cU_u$. This means that
the u-mixture is a coarsening of the t-mixture,
and the latter is thus proved to be
minimal.\hfill $\Box$

\subsection{The pure part of information gain}

We return now to our investigation of an
arbitrary bipartite state $\rho_{12}$. We have
defined $\enskip B_2^{ess}=\sum_sb_sP_2^s\enskip$
to eliminate redundant noise.

{\bf Definition 5:} We define $\enskip
C_2^{tw}\equiv \sum_tc_tP_2^t\enskip$ as a
coarsening of $B_2^{ess}$ that induces {\it
m-chaining} (cf definitions 4 and 3) of the
distant mixture $\enskip
\rho_1=\sum_sp_s\rho_1^s\enskip$ (induced by
$B_2^{ess}$), and by a spectrum $\enskip
\{c_t:\forall t\}\enskip$ consisting of any
distinct nonzero real numbers.

If $\enskip \{Q_1^t:\forall t\}\enskip$ are the
range projectors of the distant admixed states
$\enskip \rho_1^t,\enskip$ then defining,
further, $\enskip C_1^{tw}\equiv
\sum_tc'_tQ_1^t,\enskip$ ( the eigenvalues $c'_t$
any distinct and nonzero real numbers), then,
according to proposition 8 and the discussion
after its proof, one obtains {\bf twin
observables} $\enskip
(C_1^{tw},C_2^{tw}).\enskip$

{\bf Corollary 2:} In case $C_2^{tw}$ is
nontrivial, one has two parallel orthogonal
mixtures with the common index $t$, the nearby
one $\enskip \rho_2=\sum_tp_t\rho_2^t,\enskip$
and the distant one $\enskip
\rho_1=\sum_tp_t\rho_1^t.\enskip$ In general,
$\rho_{12}$ is not a mixture of the global states
$\enskip \rho_{12}^t\equiv
P_2^t\rho_{12}P_2^t/p_t,\enskip$ which give
$\rho_2^t$ and $\rho_1^t$ as their reductions.
The global states $\enskip \rho_{12}^t\enskip$
are {\it biorthogonal}, i. e., $\enskip t\not=
t'\enskip \Rightarrow \enskip
\rho_i^t\rho_i^{t'}=0,\enskip i=1,2.$

Since $C_2^{tw}$ is a coarsening of $B_2^{ess}$,
the information gain $J_{C_2^{tw}}$ of the former
is not larger than that of the latter (see
theorem 2), i. e., $$J_{C_2^{tw}}\leq
J_{B_2^{ess}}=J_{A_2},\eqno{(37a)}$$ and
$$J_{C_2^{tw}}=H(p_t)\leq H(p_s)\leq H(p_l)
\eqno{(37b)}$$ (cf (31b)). One should remember
that $\enskip H(p_s)\geq J_{A_2},\enskip$ due to
(possible) essential noise.

Observable coarsening (or o-coarsening) "$\c$" is
the opposite relation to (state-dependent)
observable refinement (or o-refinement) explained
in section II. It is a reflexive and transitive
binary relation, i. e., it is a partial order in
the set of all observables. One has
$$C_2^{tw}\c B_2^{ess}\c A_2\eqno{(37c)}$$
parallelling (37a) and (37b).\\

Returning to the elaborate subsystem entropy
decomposition (28b), and having the relations
(37a) and (37b) in mind, one can write
$$H(p_l)=S(A_2,\rho_2)=\Big\{H(p_l)-H(p_s)\Big\}+$$
$$\Big\{H(p_s)-J_{B_2^{ess}}\Big\}+
\Big\{J_{B_2^{ess}}-H(p_t)\Big\}+H(p_t).\eqno{(38)}$$

The {\it physical interpretation} of (38) goes as
follows. The entropy $\enskip
S(A_2,\rho_2)\enskip$ (quantifying the
uncertainty) of the initial subsystem observable
$A_2$ in the nearby local state $\rho_2$ consist
of redundant noise $\enskip
\big\{H(p_l)-H(p_s)\big\},\enskip$ of essential
noise $\enskip
\big\{H(p_s)-J_{B_2^{ess}}\big\},\enskip$ of {\it
garbled information gain} $\enskip
\big\{J_{B_2^{ess}}-H(p_t)\big\},\enskip$ and,
finally of {\it pure information gain} $\enskip
H(p_t).$ Naturally, each of the terms is positive
or zero, as the case may be. The latter occurs
when the corresponding subsystem observable is
trivial, i. e., when it has only one detectable
eigenvalue (the probability of which is then, of
course, one).

For different choices of $A_2$ one may come to
different $C_2^{tw}$. One may wonder if there
always exists one $C_2^{tw}$ for all choices of
$A_2$, as refined as possible. This is not true
in the general case. Taking as an example the
well-known singlet pure bipartite state, it is
easy to see that for any choice of a nontrivial
$A_2$, one has $\enskip
A_2=B_2^{ess}=C_2^{tw},\enskip$ and one obtains
thus infinitely many different $C_2^{tw}$
observables that are all complete.

The case of pure bipartite states deserves a
separate discussion (see subsection H). But first
we again need some more general theory.

\subsection{Minimal compatible coarsening of an observable}

We begin by some relevant theory on relation
between observable and state.

{\bf Definition 6:} We say that two (equal or
distinct) index values $t$ and $t'$ of {\it
detectable} eigenvalues $c_t$ and $c_{t'}$
respectively of a given observable $\enskip
C=\sum_{t''}c_{t''}P_{t''}\enskip$ are {\it
linked} with respect to a given state $\rho$ if
$\enskip P_t\rho P_{t'}\not= 0.\enskip$ When $t$
and $t'$ are such that there exists an integer
$\enskip n,\quad n=1,\enskip \mbox{or}\enskip
2,\enskip \mbox{or}\enskip \dots ,\enskip$ and
there can be found a chain of index values
$\enskip \{t_i:i=1,2,\dots ,n\}\enskip$ such that
$\enskip t=t_1,\enskip$ $\enskip t_n=t',\enskip$
and any two neighboring index values in the chain
are linked, then we say that $t$ and $t'$ are
{\it chained}. Occasionally, when it is desirable
to make a distinction with respect to m-linking
and m-chaining, we shall speak of o-linking
(short for observable-linking) and o-chaining.

O-chaining includes o-linking, and it is
reflexive, symmetric and transitive, i. e., an
equivalence relation in the detectable part of
the spectrum of the observable $C$.

{\bf Definition 7:} Let $\enskip
C=\sum_tc_tP_t\enskip$ be a given observable, and
let the observable $D$ be a {\it coarsening} of
$C$ defined by means of {\it chaining} in the
detectable part of the spectrum of the latter
with respect to a given state $\rho$ (cf
definition 6):
$$D\equiv \sum_kd_kP_k,\quad \forall k:\enskip
P_k\equiv \sum_{t\in \cC_k}P_t,\eqno{(39a)}$$
where
$$\{t:\forall t,p_t\equiv \tr (\rho P_t)>0\}=
\sum_k\cC_k\eqno{(39b)}$$ partitions the
detectable part of the spectrum of $C$ into the
equivalence classes $\cC_k$ obtained by
o-chaining, and the eigenvalues of $D$ are
arbitrary distinct nonzero real numbers. We call
$D$ the {\it chained coarsening} of $C$ with
respect to $\rho$.

{\bf Proposition 10:} The chained coarsening
$\enskip D\enskip \big(\co C\big)\enskip$ given
in definition 7
 is {\it compatible} with $\rho$,
$\enskip [D,\rho ]=0\enskip$. It is {\it the most
refined} coarsening of $\enskip
C=\sum_tc_tP_t\enskip$ compatible with $\rho$, i.
e., if $\enskip \bar C\equiv \sum_j\bar c_j\bar
P_j,\enskip$ is a coarsening of $C$, $\enskip
\forall j:\enskip \bar P_j\equiv \sum_{t\in \bar
\cC_j}P_t\enskip$ with arbitrary distinct nonzero
real eigenvalues of $\bar C$ and $\enskip
\{t:\forall t, p_t>0\} =\sum_j\bar \cC_j\enskip$
a partitioning of the detectable part of the
spectrum of $C$, such that $\bar C$ is compatible
with $\rho$, $\enskip \bar C\co C,\enskip$ $[\bar
C,\rho ]=0,\enskip$ then it is also a coarsening
of $D$: $\enskip \bar C\co D.$

{\bf Proof:} Let $k,k'$ be two distinct index
values of $D$, and let us keep in mind that
$\enskip P_k\rho P_{k'}=\Big(\sum_{t\in
\cC_k}P_t\Big)\rho \Big(\sum_{t'\in
\cC_{k'}}P_{t'}\Big).\enskip$ Since $t$ and $t'$
are not o-chained by assumption, they are not
o-linked either. Hence, each term is zero
$\enskip P_t \rho P_{t'}=0\enskip$ (cf definition
6), implying $\enskip P_k\rho P_{k'}=0.\enskip$
Thus, one can write $\enskip \rho
=\Big(\sum_kP_k\Big)\rho \Big(\sum_{k'}
P_{k'}\Big)=\sum_k(P_k\rho P_k),\enskip$ i. e.,
$\rho$ is compatible with each eigenprojector of
$D$, hence also with $D$ itself.

Let $\bar C$ be an arbitrary coarsening of $C$
compatible with $\rho$ as given in the theorem.
Let $\enskip t\in \bar \cC_j,\enskip t'\in \bar
\cC_{j'}.\enskip$ Since $\enskip j\not= j'\enskip
\Rightarrow \enskip \bar P_j\rho \bar P_{j'}=0,
\enskip$ due to $\enskip [\bar C,\rho
]=0,\enskip$ multiplication from the left by
$P_t$ and from the right by $P_{t'}$ gives
$\enskip P_t\rho P_{t'}=0,\enskip$ i. e., $t,t'$
are not o-linked. Equivalently, if $t,t'$ are
o-linked, then $\enskip j=j',\enskip$. In other
words, o-linked index values, and hence also
o-chained index values, belong to one and the
same equivalence class $\bar \cC_j$. Thus,
$\enskip \forall k:\enskip \exists j:\enskip
\cC_k\subseteq \bar \cC_j.$ This proves $\enskip
\bar C\co D.$\hfill $\Box$

\subsection{Twin observables for pure bipartite
states}

As it was stated, if $(C_1,C_2)$ are twin
observables in any bipartite state $\rho_{12}$,
we have the necessary local condition of
compatibility of observable and state
$[C_2,\rho_2]=0$ (and symmetrically,
$[C_1,\rho_1]=0$). If one deals with a pure state
$\enskip
\rho_{12}=\ket{\Phi}_{12}\bra{\Phi}_{12},\enskip$
then this condition is {\it also sufficient}
\cite{FHMV}, \cite{MVFH}.

If the bipartite state is pure, then for finding
$C_2^{tw}$ for a given local observable
$B_2^{ess}$ one need not resort to the distant
mixture $\enskip \rho_1=\sum_sp_s\rho_1^s$.

{\bf Theorem 4:} If the bipartite state is {\it
pure}, and a nearby local observable $\enskip
A_2= \sum_la_lP_2^l\enskip$ is given, then the
observable $C_2^{tw}$ (cf subsection F) is the
chained coarsening of $A_2$ with respect to
$\rho_2$ (cf definitions 6 and 7).

{\bf Proof:} Proposition 10 makes it clear that
those and only those coarsenings of a given
observable with respect to a given state are
compatible with the state that are coarsenings of
the chained coarsening of the observable.
Naturally, the latter is the most refined one.
That is precisely what $C_2^{tw}$ is regarding
$A_2$.\hfill $\Box$

At first glance one might wonder why is
$C_2^{tw}$ not the chained coarsening of $A_2$
with respect to $\rho_2$ for a general state
$\rho_{12}$. The answer lies, of course, in the
fact that there may exist coarsenings
$C_2=\sum_tp_tP_2^t$ of $A_2$ that are compatible
with $\rho_2$ and that do not induce orthogonal
distant mixtures $\enskip \rho_1=
\sum_tp_t\rho_1^t$. This is so because
compatibility of local observable with local
state is in general, in contrast to the special,
pure-state case, not sufficient for twin
observables. Note that orthogonality of both
mixtures $\enskip \rho_i=
\sum_tp_t\rho_i^t,\enskip i=1,2\enskip$ is a
characteristic property of twin observables (cf
the "measurement-theoretic condition" $\enskip
\forall t:
P_i^t\rho_{12}P_i^t=P_{i'}^t\rho_{12}P_{i'}^t,
\enskip i\not= i',\enskip i,i'=1,2\enskip$
defining twin observables in \cite{FHPR02}).

Let a subsystem observable be complete $\enskip
A_2^c=\sum_la_l\ket{l}_2\bra{l}_2,\enskip$ and
let
$$\ket{\Phi}_{12}=
\sum_l'\alpha_l\ket{l}_1\otimes
\ket{l}_2\eqno{(40b)}$$ be a (generalized)
expansion of $\ket{\Phi}_{12}$ in the eigenbasis
$\enskip \{\ket{l}_2:\forall l\}\enskip$ of the
subsystem observable $A_2$, omitting undetectable
index values, and the $\ket{l}_1$ being unit
vectors. (It is non-unique because the phase
factors of the $\ket{l}_1$ vectors are not
specified.) Then (40b) implies that
$$\forall l:\quad p_l=|\alpha_l|^2,\quad
\rho_1^l=\ket{l}_1\bra{l}_1$$ in (40a) as easily
seen.

Since for a pure state $\ket{\Phi}_{12}$
compatibility of local observable and local state
is necessary and sufficient for possessing an
(opposite-subsystem) twin observable, one can
distinguish two kinds of choices for $A_2$. In
the first, $A_2$ is incompatible with $\rho_2$.
Then, if there exists a nontrivial orthogonal
decomposition of the range of $\rho_2$ that is
invariant both for $A_2$ and for $\rho_2$, then
there exists a finest of this kind defining a
nontrivial $C_2^{tw}$. In the second choice,
$A_2$ is compatible with $\rho_2$. Then $\enskip
B_2^{ess}=\sum_l'b_lP_2^l,\enskip$ and $\enskip
C_2^{tw}=\sum_l'c_lP_2^L,\enskip$ where the prim
denotes restriction to detectable eigenvalues
$a_l$ of $A_2$ (and the eigenvalues are distinct,
nonzero, arbitrary real numbers).

If in the second choice $A_2$ is complete
$\enskip
A_2^c=\sum_la_l\ket{l}_2\bra{l}_2,\enskip$ then
(40b) is the well-known {\it Schmidt expansion}
if, additionally, the phase factors of
$\ket{l}_1$ are chosen so that $\alpha_l$ are
positive. Then $\enskip \forall l, p_l>0:\enskip
\alpha_l=r_l^{1/2},\enskip$ and $\enskip \rho_i=
\sum_l'r_l\ket{l}_i\bra{l}_i,\enskip
i=1,2,\enskip$ are spectral forms of the
reductions. One has a Schmidt expansion if the
(generalized) expansion is in the eigenbasis of
one reduction, and only if it is in those of both
reductions (and if the numerical expansion
coefficients are positive) \cite{FHMV}. (One
should note that the entire non-uniqueness of a
Schmidt expansion is in the
choice of an eigenbasis of one reduction.)\\

The string of inequalities (37c) can be continued
in the general case. Namely, $\enskip
I_C(C_2^{tw},\rho_2)=0\enskip$. Thus, $\rho_2$
has no quantumness with respect to $C_2^{tw}$,
but, in general, there still is quantumness in
the mutual information in the form of discord
$\enskip
\delta_{C_2}=I_C(C_2,\rho_{12})>0.\enskip$ The
next and last step to be taken is to eliminate
also this (possible) quantumness.

\subsection{Quasi-classical correlations}

Let us now return to the general case of a
bipartite state $\rho_{12}$, and the initial
subsystem observable $\enskip
A_2=\sum_la_lP_2^l.$

{\bf Definition 8:} Let the observable $\enskip
D_2^{qc}\equiv \sum_kd_k P_2^k\enskip$ be the
chained coarsening of $C_2^{tw}$ with respect to
$\rho_{12}$ (with distinct nonzero real
eigenvalues).

{\bf Theorem 5:} The following strings of
inequalities are valid:
$$D_2^{qc}\c C_2^{tw}\c B_2^{ess}\c A_2.
\eqno{(41a)}$$
$$J_{D_2^{qc}}\leq J_{C_2^{tw}}\leq
J_{B_2^{ess}}=J_{A_2},\eqno{(41b)}$$
$$J_{D_2^{qc}}=H(p_k)\leq J_{C_2^{tw}}=H(p_t)\leq
H(p_s)\leq H(p_l)\eqno{(41c)}$$

{\bf Proof:} The first "inequality" in (41a)
follows directly from definition 8, and then, due
to transitivity, the rest of them are
consequences. The rest of inequalities in theorem
5 are implied by theorem 2 and (31b).\hfill
$\Box$\\

One can write
$$H(p_l)=S(A_2,\rho_2)=\big\{H(p_l)-H(p_s)\big\}+
\big\{H(p_s)-J_{B_2^{ess}}\big\}+$$ $$
\big\{J_{B_2^{ess}}-H(p_t)\big\}+\{H(p_t)-H(p_k)\}+H(p_k).
\eqno{(42)}$$

The last two terms in (42) are possibly positive
pure information gain. It consists of a purely
quantum term $\enskip \{H(p_t)-H(p_k)\},\enskip$
and a {\it quasi-classical} term $\enskip
H(p_k).\enskip$

{\bf Corollary 3:} On account of the
compatibility $\enskip
[D_2^{qc},\rho_{12}]=0\enskip$, the subsystem
observable $D_2^{qc}$, if non-trivial, induces
not only the subsystem mixtures $\enskip
\rho_i=\sum_kp_k\rho_i^k,\enskip i=1,2,\enskip$
but also the {\it global mixture} $$
\rho_{12}=\sum_k p_k\rho_{12}^k,\eqno{(43)}$$
where $\enskip \forall k:\enskip
\rho_{12}^k\equiv P_2^k\rho_{12}/p_k\enskip$. The
mixture is {\it biorthogonal}, i. e., $\enskip
k\not= k'\enskip \Rightarrow \enskip
\rho_i^k\rho_i^{k'}=0,\enskip i=1,2.$

{\bf Proof:} The claimed compatibility of
$D_2^{qc}$ and $\rho_{12}$ follows from
definition 8 and proposition 10. It implies
compatibility of $D_2^{qc}$ and $\rho_2$. In
general, when $\enskip A_2=\sum_la_lP_2^l\enskip$
is given, the corresponding nearby mixture is
$\enskip \rho_2^L(A_2)\equiv
\sum_lP_2^l\rho_2P_2^l,\enskip$ which, in this
case amounts to the orthogonal mixture $\enskip
\rho_2=\sum_kp_k(P_2^k\rho_2/p_k).\enskip$

Orthogonality of the distant mixture is implied
by the first "inequality" in (41a): any
coarsening of an orthogonal mixture is
orthogonal. The global mixture (43) itself is, of
course, due to $[D_2^{qc},\rho_{12}]=0$.\hfill
$\Box$\\

When one writes down the elaborate subsystem
entropy decomposition (28b) for $D_2^{qc}$
(changing what has to be changed), then $\enskip
0=I_C(D_2^{qc},\rho_{12})=I_C(D_2^{qc},\rho_2)=
\delta_{D_2^{qc}}. \enskip$ Thus, if we
"interrogate" $\rho_{12}$ by $D_2^{qc}$, then
quantumness has disappeared not only in the
nearby subsystem (analogously as due to
$C_2^{tw}$), but also in the global system, i.
e., it has disappeared completely. For this
reason we call this last step "quasi-classical".

\subsection{Measures of quantumness}

We now complete inequalities (41b), (41c), and
equality (42) by analogous relations for the
measures of quantumness.

{\bf Theorem 6: A)} The following strings of
inequalities parallel (41a): $$I_C(D_2^{qc},
\rho_{12})\leq I_C(C_2^{tw},\rho_{12})\leq
I_C(B_2^{ess},\rho_{12})\leq
I_C(A_2,\rho_{12});$$ $$I_C(D_2^{qc},\rho_2) \leq
I_C(C_2^{tw},\rho_2)\leq I_C(B_2^{ess},\rho_2)
\leq I_C(A_2,\rho_2);$$ $$\delta_{D_2^{qc}}\leq
\delta_{C_2^{tw}}\leq \delta_{B_2^{ess}}\leq
\delta_{A_2}.$$

{\bf B)} The coherence informations satisfy also
the straight-line relations: $$I_C(A_2,\rho_i)=
I_C(D_2^{qc},\rho_i)+I_C(C_2^{tw},\sum_kP_2^k\rho_i
P_2^k)+$$ $$
I_C(B_2^{ess},\sum_t\sum_kP_2^tP_2^k\rho_i
P_2^kP_2^t)+$$ $$I_C(A_2,\sum_s\sum_t\sum_k
P_2^sP_2^tP_2^k\rho_iP_2^kP_2^tP_2^s),\quad
i=2,12.$$

{\bf Proof: A)} The first two strings of
inequalities are an immediate consequence of the
inequality proved in previous work \cite{Roleof}
(Theorem 3 there, "$E_C$" is written instead of
"$I_C$"). The third string of inequalities is an
immediate consequence of theorem 2.

{\bf B)} The straight-line relations are an
immediate implication of Corollary 2 in
\cite{FHJP05}.\hfill $\Box$

\section{Examples}

\subsection{Pure states}

{\bf Example 1:} Let $\enskip
\{\ket{i}_2:i=1,2,3\} \enskip$ be an orthonormal
set in the state space of the nearby subsystem,
and let $\enskip \{\ket{j}_1:j=1,2\}\enskip$ be
an orthonormal set in that of the distant
subsystem. We define
$$\ket{\Phi}_{12}\equiv
\alpha_1\ket{j=1}_1\ket{i=1}_2+
\alpha_2\ket{j=1}_1\ket{i=2}_2+
\alpha_3\ket{j=2}_1\otimes \ket{i=3}_2,
\eqno{(44a)}$$ where, of course, $\enskip
\sum_{q=1}^3|\alpha_q|^2=1\enskip$ is valid.

If $\enskip \{\ket{i}_2:i=1,2,3\}\enskip$ is a
subset of the eigenbasis of a complete subsystem
observable $A_2$, then the corresponding distant
state decomposition is
$$\rho_1=|\alpha_1|^2\ket{j=1}_1
\bra{j=1}_1+|\alpha_2|^2\ket{j=1}_1\bra{j=1}_1+
|\alpha_3|^2\ket{j=2}_1\bra{j=2}_1$$ (cf (40b)
and the next relation). It contains repetition in
the admixed states. This makes the subsystem
observables $\enskip B_2^{ess}=C_2^{tw}\equiv
b_1P_2^{s=1}+b_2\ket{s=2}_2\bra{s=2}_2, \enskip$
where $\enskip P_2^{s=1}\equiv
\ket{i=1}_2\bra{i=1}_2+\ket{i=2}_2\bra{i=2}_2
\enskip$ and $\enskip \ket{s=2}_2\equiv
\ket{i=3}_2,\enskip$ nontrivial. The
corresponding pure information gain is
$$\enskip J_{B_2^{ess}}=H(p_{s=1},
p_{s=2})=S(B_2^{ess},\rho_2)=S(C_2^{tw},\rho_2)=$$
$$-(|\alpha_1|^2+|\alpha_2|^2)log(|\alpha_1|^2+
|\alpha_2|^2)-|\alpha_3|^2log|\alpha_3|^2.$$

A  Schmidt expansion of $\ket{\Phi}_{12}$ is
$$\ket{\Phi}_{12}=r_1^{1/2}\ket{j=1}_1\ket{r_1}_2+
|\alpha_3|\big(e^{i\lambda_3}\ket{j=2}_1\big)
\ket{i=3}_2,\eqno{(44b)}$$ where $\enskip
r_1=|\alpha_1|^2+|\alpha_2|^2,\enskip$
$e^{i\lambda_3}$ is the phase factor of
$\alpha_3$, and
$$\ket{r_1}_2\equiv (\alpha_1\ket{i=1}_2+
\alpha_2\ket{i=2}_2)/r_1^{1/2}.\eqno{(44c)}$$

{\bf Example 2:} We assume that all positive
eigenvalues of $\rho_2$ of $\ket{\Psi}_{12}$ are
{\it non-degenerate}. Let $\enskip
\{\ket{q}_2:\forall q\}\enskip$ be the unique (up
to phase factors) eigen-sub-basis of $\rho_2$
corresponding to its positive eigenvalues
$\enskip \{r_q>0:\forall q \}.\enskip$ Finally,
let $$
\ket{\Psi}_{12}=\sum_qr_q^{1/2}\ket{q}_1\otimes
\ket{q}_2\eqno{(45a)}$$ be a Schmidt expansion of
$\ket{\Psi}_{12}$.

Let $A_2^c=\sum_l\ket{l}_2\bra{l}_2$ be a
complete observable. Then the induced nearby
mixture is
$$\rho_2^L(A_2^c)\equiv \sum_lP_2^l\rho_2P_2^l=
\sum_lp_l\ket{l}_2\bra{l}_2.\eqno{(45b)}$$

Let the eigenbasis of $A_2^c$ contain $\enskip
\{\ket{q}_2:\forall q\}\enskip$ as a subset.
Then, as easily seen from theorem 4, $\enskip
B_2^{ess}= \sum_qb_q\ket{q}_2\bra{q}_2,\enskip$
$\enskip
C_2^{tw}=\sum_qc_q\ket{q}_2\bra{q}_2,\enskip$
$\enskip
Q_2A_2^c=\sum_qa_q\ket{q}_2\bra{q}_2,\enskip$ and
$\enskip D_2^{qc}=Q_2,\enskip$ where $Q_2$ is the
range projector of $\rho_2$. (The observable
$D_2^{qc}$ is trivial because a pure state cannot
be written as a nontrivial mixture - cf corollary
3.)

Let us now take another complete observable
$A_2^c$ as follows. Let $\enskip
\ket{l=1}_2\equiv \ket{q=1}_2\enskip$ from above.
Further, let $\enskip
\bra{l}_2\ket{q}_2\bra{q}_2\ket{l'}_2 \not=
0\enskip$ unless $\enskip q=1,\enskip$ and at
least one of the index values $l,l'$ is not equal
to $1$, when it is zero. Then, as easily seen,
theorem 4 implies that $\enskip C_2^{tw}=
c_1\ket{q=l=1}_2\bra{q=l=1}_2+c_2P_2^{t=2},\enskip$
where $\enskip P_2^{t=2}\equiv \sum_{q\geq 2}
\ket{q}_2\bra{q}_2.$

\subsection{Mixed states}

{\bf Example 3:} We assume that all vectors
$\ket{q}_1$ in (45a) are orthogonal to all
$\ket{j}_1$ in (44a), and symmetrically, that all
$\ket{q}_2$ in (45) are orthogonal to all
$\ket{i}_2$ in (44a). Then we take a mixture of
the bipartite pure state vectors given by (44b)
and (45a): $\enskip \rho_{12}\equiv (1/2)
\ket{\Phi}_{12}\bra{\Phi}_{12}+
(1/2)\ket{\Psi}_{12}\bra{\Psi}_{12}.\enskip$

Further, we define $A_2$ to be complete and such
that its eigenbasis contains all the mentioned
orthonormal vectors for the nearby subsystem as
subsets. Then all four observables in (41a) are
nontrivial: $\enskip
D_2^{qc}=d_{k=1}P_2^{k=1}+d_2P_2^{k=2},\enskip$
where $\enskip P_2^{k=1}\equiv
\sum_{i=1}^3\ket{i}_2\bra{i}_2$, and $\enskip
P_2^{k=2}\equiv \sum_q\ket{q}_2 \bra{q}_2$;
$$C_2^{tw}=B_2^{ess} \equiv
c_{t=1}\ket{r_1}_2\bra{r_1}_2+
c_{t=2}\ket{i=3}_2\bra{i=3}_2+$$ $$
\sum_qc_{t=q+3}\ket{t=q+3}_2\bra{t=q+3}_2$$ (cf
(44c)).

\section{Has the Discord Disappeared in Measurement?}

As it was stated, to extract the information gain
$J_{A_2}$ from $\rho_{12}$, one measures $A_2$
locally on the nearby subsystem, and by this very
fact also $(1\otimes A_2)$ globally in
$\rho_{12}$. In general, one thus obtains
$\enskip S(A_2,\rho_2)=H(p_l),\enskip$ in which
to $J_{A_2}$ is inseparably added both the
essential and the redundant noise, and $J_{A_2}$
necessarily contains garbled information gain in
the general case. If it contains a positive
amount of pure information, this, in turn,
consists of a quantum and a quasi-classical term.

As far as quantities are concerned, the results
of the preceding section allow one to evaluate
how much of each of the mentioned entities is
contained in $H(p_l)$. But qualitatively, when
one deals with an ensemble $\rho_{12}$ of
individual bipartite systems in the laboratory,
on each of which $A_2$ is measured, at first
glance, one can do nothing in the way of
separation of these entities. One can, of course,
measure locally $B_2^{ess}$ (or $C_2^{tw}$ or
$D_2^{qc}$) instead of $A_2$. Actually, if the
laboratory ensemble is sufficiently large, the
thing to do is to measure the mentioned
observables on subensembles, which, if randomly
taken, also represent empirically the same
bipartite state $\rho_{12}$.

The simplest way to measure $A_2$ is the
so-called ideal measurement, which, by
definition, changes a state $\rho$ into its
L\"{u}ders mixture $\rho_L$ \cite{Lud}. Then
relations (5) and (4a) imply  $\enskip
I(\rho_{12}^L)=I(\rho_{12})-\delta_{A_2}.
\enskip$ The discord has disappeared from the
bipartite state. Hence the title of this section.

In addition to this disappearance, one has the
following known fact.

{\bf Lemma 7:} If $\rho_{12}^f\equiv (U_1\otimes
U_2)\rho_{12}(U_1\otimes U_2)^{\dagger}$,where
$\rho_{12}$ is an arbitrary bipartite state,
$U_i,\enskip i=1,2$ are any unitary subsystem
operators, and the suffix {f} denotes "final",
then
$$I(\rho_{12}^f)=I(\rho_{12}).$$

Putting it in words, in any bipartite
state, when it is dynamically closed and
the two subsystems do not interact, {\it
the mutual information does not change}.

{\bf Proof} is straightforward.

To apply Lemma 7 to the case of {\it ideal
measurement} of $A_2$ in $\rho_{12}$, let the
instrument that performs a measurement of the
observable be subsystem $3$. Subsystems $1$ and
$(2+3)$ do not interact during the subsystem
measurement, and the tripartite system is
dynamically closed. Hence, according to Lemma 7,
the mutual information between subsystems $1$ and
$(2+3)$ does not change. Writing $\rho_{1,23}^f$
for the state $\rho_{123}^f$ of the bipartite
system $1+(2+3)$, we have
$$I(\rho_{1,23}^i)=I(\rho_{1,23}^f)$$ (the suffix
"$i$" denotes "initial"). Further, strong
subadditivity of entropy requires that
$I_{12}\leq I_{1,23}$ be always valid (cf
relation (7) in \cite{Mutual}). On the other
hand, the initial state $\rho_3^i$ of the
measuring apparatus is uncorrelated with the
$(1+2)$ system at the beginning of measurement,
i. e., $\rho_{123}^i\equiv \rho_{12}\otimes
\rho_3^i$. Hence we have a case of equality in
the strong subadditivity of entropy inequality:
$$I(\rho_{12})=I(\rho_{1,23}^i)$$
\cite{Mutual} (see relation (8) there).
Altogether,
$$I(\rho_{12})=I(\rho_{1,23}^f).\eqno{(46a)}$$

Thus, the amount of mutual information
between subsystems $1$ and $2$ at the
beginning of measurement is {\it
preserved} as the amount of mutual
information between subsystems $1$ and
$(2+3)$ at the end of ideal measurement.

Strong subadditivity of entropy requires $\enskip
I(\rho_{12}^f)\leq I(\rho_{1,23}^f)\enskip$,
where $\enskip \rho_{12}^f\equiv
\tr_3\rho_{123}^f,\enskip$ and, in case of ideal
measurement, it is seen from (4a) and (5) that we
now have a proper inequality: $I(\rho_{12}^f)<
I(\rho_{1,23}^f) =I(\rho_{12})$ in the general
case.

The final mutual information $I(\rho_{1,23}^f)$
can be decomposed according to (4a) (changing
what has to be changed) with respect to the same
observable $A_2$:
$$I(\rho_{1,23}^f)=\sum_l\Big(p_l^fS(\rho_1^{fl}
||\rho_1^f)\Big)+ \Big(I_C(A_2,\rho_{123}^f)
-I_C(A_2,\rho_{23}^f)\Big)+\sum_l\Big(p_l^f
I(\rho_{1,23}^{fl})\Big),\eqno{(46b)}$$ where the
suffix $f$ denotes that the quantity is derived
from the final state $\rho_{123}^f$, and $l$
stems from the eigenprojector $P_2^l$ of $A_2$.
In particular, $\forall l:\enskip p_l^f\equiv \tr
(\rho_{123}^fP_2^l)$; $$\forall l,\enskip
p_l^f>0:\enskip \rho_{1}^{fl}\equiv
\tr_{23}(\rho_{123}^f
P_2^l/p_l^f)=\tr_{23}(P_2^l\rho_{123}^fP_2^l
/p_l^f);$$ etc.

To find out how each of the three terms changes
from $\rho_{12}$ to $\rho_{1,23}^f$, i. e., from
(4a) to (46b), we define {\it the simplest
measuring apparatus} for ideal measurement:

The initial state of subsystem $3$ is pure
$\rho_3^i\equiv \ket{\phi}_3\bra{\phi}_3$; the
"pointer observable" is a complete one
$A_3=\sum_lb_l \ket{l}_3\bra{l}_3$ (spectral form
in terms of distinct eigenvalues - "pointer
positions"); finally, the interaction evolution
goes as follows $$\rho_{123}^f=
U_{23}(\rho_{12}\otimes
\ket{\phi}_3\bra{\phi}_3)U_{23}^{\dagger},
\eqno{(47a)}$$ and it is such that
$$\forall l:\quad \tr (\rho_{123}^f
\ket{l}_3\bra{l}_3)=p_l\eqno{(47b)}$$ (cf (2a)),
and $$\forall l,\enskip p_l>0:\quad
p_l^{-1}\tr_3(\rho_{123}^f
\ket{l}_3\bra{l}_3)=\rho_{12}^l \eqno{(47c)}$$
(cf (2b)).

{\bf Theorem 7:} Comparing (4a) and (46b), {\bf
all} corresponding quantities on the RHSs are
{\it equal}. More precisely,
$$\forall l:\quad
p_l=p_l^f,\eqno{(48a)}$$
$$\forall l,\enskip p_l>0:\quad
\rho_1^l=\rho_1^{fl},\eqno{(48b)}$$
$$\rho_1=\rho_1^f,\eqno{(48c)}$$
$$I_C(A_2,\rho_{12})=I_C(A_2,\rho_{123}^f),
\eqno{(48d)}$$ $$I_C(A_2,\rho_2)=
I_C(A_2,\rho_{23}^f),\eqno{(48e)}$$ where
$\enskip \rho_{23}^f\equiv
\tr_1\rho_{123}^f,\enskip$ and finally
$$\forall l,\enskip p_l>0:\quad
I(\rho_{12}^l)=I(\rho_{1,23}^{fl}).\eqno{(48f)}$$

The theorem is proved in Appendix C.\\

For further use, we establish that $A_2$ and
$A_3$ are {\it twin observables} in relation to
$\rho_{12,3}^f$. Subsystem $3$ is viewed as the
nearby one, and the bipartite system $(1+2)$ as
the distant one.

As it was shown in subsection IV.D and
proposition 8, it is sufficient to point out that
$\enskip \sum_l\ket{l}_3\bra{l}_3=1,\enskip$ and
$\enskip
(\sum_l'\ket{l}_3\bra{l}_3)\rho_3^f=\rho_3^f\enskip$
(the undetectable $l$ values are omitted). It
follows from (47c) that $\enskip
\sum_l'p_l\rho_{12}^l=\rho_{12}^f\equiv
\tr_3\rho_{123}^f,\enskip$ with all weights
positive and the admixed states $\rho_{12}^l$
orthogonal because $\enskip
\rho_{12}^l=P_2^l\rho_{12}^lP_2^l\enskip$ (cf
(2b)). Hence, according to proposition 8, $A_2$
and $A_3$ are twin observables in
$\rho_{12,3}^f$.\\

It is also of interest to consider a relevant
subsystem entropy decomposition for
$\rho_{12,3}^f$, i. e., for the bipartite system
$(1+2)+3$ in the state $\rho_{123}^f$.

{\bf Theorem 8:} The following subsystem entropy
decomposition in terms of $A_2$ and $\rho_{12}$
entities is valid:
$$S(\rho_{123}^f)=S(\rho_{12}^f)-I(\rho_{12,3}^f)+
S(\rho_3^f)=$$
$$\Big\{S(\rho_{12})+I_C(A_2,\rho_{12})\Big\}-
\Big\{I_C(A_2,\rho_{12})+H(p_l)\Big\}+\Big\{H(p_l)\Big\}.
\eqno{(49)}$$ (It is understood that each
large-brackets expression equals the
corresponding entity in the preceding
decomposition.)

{\bf Proof:} Relation (47c) implies
$$\rho_{12}^f\equiv
\tr_3(\rho_{123}^f)=\sum_l\Big(\tr_3(\rho_{123}^f
\ket{l}_3\bra{l}_3)\Big)=\sum_lp_l\rho_{12}^l.$$
Hence, $$\rho_{12}^f=\rho_{12}^L, \eqno{(50)}$$
i. e., it is the L\"{u}ders mixture of the
initial state $\rho_{12}$ with respect to $A_2$.
Further, definition (1a) gives $\enskip
I_C(A_2,\rho_{12})\equiv S(\rho_{12}^L)
-S(\rho_{12}).\enskip$ Thus, the first
large-brackets expression follows.

Next we prove the third large-brackets
expression. The pointer observable $A_3$, being a
twin observable, is necessarily compatible with
$\rho_3^f$. Since it is also complete (by
definition), its entropy coincides with the
entropy of $\rho_3^f$: $\enskip
S(A_3,\rho_3^f)=H(p_l)= S(\rho_3^f)$.

Finally, in view of the fact that the system
$(1+2+3)$ is dynamically closed (isolated) during
the measurement interaction, the total entropy is
preserved: $\enskip S(\rho_{123}^f)=
S(\rho_{12}\otimes \ket{\phi }_3\bra{\phi}_3)=
S(\rho_{12}).\enskip$ The second large-brackets
expression follows from this.\hfill $\Box$\\

We have to clarify how theorem 8 relates to the
proved disappearance of the discord $\enskip
\delta_{A_2}(\rho_{12})=I_C(A_2,\rho_{12})
-I_C(A_2,\rho_2)\enskip$ (cf (7)) in the
measurement interaction. Since
$\delta_{A_2}(\rho_{12})$ is a term in the mutual
information (cf (28b)), at first glance one would
expect that $S(\rho_{12})$ increases by
$\delta_{A_2}(\rho_{12})$ when $\rho_{12}$ goes
over into $\rho_{12}^f$. But this is not so
because, as seen in (28b), $I_C(A_2,\rho_2)$
actually cancels out in $S(\rho_{12})$. In
$\rho_{12,3}^f$ the measured observable $A_2$ and
the pointer observable $A_3$ are twin
observables, and, as a consequence (cf (36a)),
one has compatibility $\enskip
[A_2,\rho_{12}^f]=0,\enskip [A_3,\rho_3^f]=0.
\enskip$ Therefore, we can forget about the
quantumness of $A_2$ in relation to
$\rho_{12}^f$, and we do cancel $I_C(A_2,\rho_2)$
in $I(\rho_{12})$ and $S(\rho_2)$ in (28b). Thus,
the increase in $S(\rho_{12})$ is
$I_C(A_2,\rho_{12})$ (cf (7)) in accordance with
(49).

Let us write down next (28b) for
$S(\rho_{12,3}^f)$ with respect to $A_3$:
$$S(\rho_{12,3}^f)=S(\rho_{12}^f)-I(\rho_{12,3}^f)+
S(\rho_3^f)=$$
$$\Big\{\sum_lp_lS(\rho_{12}^l)+H(p_l)\Big\}
-\Big\{H(p_l)+I_C(A_3,\rho_{123}^f)\Big\}+ \Big\{
H(p_l)\Big\}.  \eqno{(51)}$$ (One should note
that $I_C(A_3,\rho_3^f)=0$, and that on account
of $A_3$ being complete, the respective residual
terms in $I_{12,3}$ and $S(\rho_3^f)$ are zero.)

Comparing (49) and (51), one infers that
$$I_C(A_2,\rho_{12})=I_C(A_3,\rho_{123}^f)=
I_C(A_2, \rho_{123}^f).$$ The last equality is a
general property of twin observables: they have
the same coherence information in the bipartite
state as follows from (in our case) $\enskip
\forall l:\enskip
P_2^l\rho_{123}^f=\ket{l}_3\bra{l}_3\rho_{123}^f,
\enskip$ (cf (34b)), which is one of the
equivalent definitions of twin observables
\cite{FHPR02}.

Thus, we have proved

{\bf Theorem 9:} It is not the discord $\enskip
\delta_{A_2}(\rho_{12})=I_C(A_2,\rho_{12})
-I_C(A_2,\rho_2)\enskip$ (cf (7)), but only the
non-negative global term in it that is {\it
preserved} in the measurement interaction:
$$I_C(A_2,\rho_{12})=I_C(A_2,\rho_{123}^f).
\eqno{(52)}$$

{\bf Corollary 4:} One has $\enskip
A_3=C_3^{tw}\enskip$ with respect to
$\rho_{12,3}^f$, i. e., $\enskip S(A_3,
\rho_3^f)=H(p_l)\enskip$ is pure information on
the distant mixture $\enskip
\rho_{12}^f=\sum_lp_l\rho_{12}^l\enskip$ (cf
(50)), which is orthogonal.

This pure information is not the information at
issue. The subject of our investigation is
$J_{A_2}(\rho_{12})$, the information gain in the
distant mixture $\enskip \rho_1=\sum_lp_l\rho_1^l
\enskip$ induced by $A_2$ in $\rho_{12}$. Thus,
one should view $H(p_l)$ decomposed according to
(42), which shows that it consists of a
redundant-noise term $\enskip
\Big(H(p_l)-H(p_s)\Big),\enskip$ an
essential-noise term $\enskip
\Big(H(p_s)-J_{B_2^{ess}}\Big),\enskip$ a term
$\enskip \Big(J_{B_2^{ess}}-H(p_t)\Big)\enskip$
of garbled information (due to the overlap in the
admixed states $\rho_1^l$), of a term $\enskip
\Big(H(p_t)-H(p_k)\Big)\enskip$ of pure quantum
information, and, finally, of a term $H(p_k)$ of
pure quasi-classical information. Naturally, any
of these terms can be zero.

The measurement interaction, or pre-measurement
as it is called in the thorough measurement
theory \cite{BLM}, is not the final step in
measurement. It is collapse, objectification or
reduction (cf also \cite{Auletta}, which makes
ample use of \cite{B-K}), which turns
$\rho_{123}^f$ into the L\"{u}ders mixture
$$\sum_lp_l\Big(\rho_{12}^l\otimes
\ket{l}_3\bra{l}_3\Big)\eqno{(53)}$$ of
$\rho_{123}^f$ with respect to $A_3$. The admixed
L\"{u}ders states $\enskip (\rho_{12}^l\otimes
\ket{l}_3\bra{l}_3) \enskip$ correspond to the
individual results $a_l$ of $A_2$ revealed by the
pointer position $\enskip
\ket{l}_3\bra{l}_3\enskip$ of the pointer
observable $A_3$. Incidentally, the state (53) is
a quasi-classical mixture, well known in
laboratory measurements.

Both in the final state of premeasurement
$\rho_{123}^f$ and in the final state of
measurement given by (53) there are, in general,
correlations in the subsystem $(1+3)$ though $1$
and $3$ have not interacted. Thus, subsystem $1$
has simultaneous correlations with $2$ and with
$3$, and so-called monogamy \cite{Koashi},
expressing mutual restrictions in the two
mentioned correlations, enters the scene. Koashi
and Winter have quantified monogamy
\cite{Koashi}. In one of their inequalities
appears, as a measure of correlations, the
so-called entanglement of formation expressing
the least expected entanglement of any ensemble
of pure states realizing a given bipartite state
\cite{Bennett2}. Their inequality (6) can easily
be rearranged to the effect that for $(1+3)$ the
entanglement of formation {\it cannot exceed} the
minimal residual entropy $inf_{\forall
B_2}\sum_kp_kS(\rho_1^k)$ (cf (6) in this article
with $k$ instead of $l$). The latter quantity
applies to subsystem $1$ when all imaginable
choices of the observable $B_2=\sum_kb_kP_k$ (all
$b_k$ distinct eigenvalues) with a view to be
measured in the state
$\rho_{12}^f=\sum_lp_l\rho_{12}^l$ of subsystem
$(1+2)$ are taken into account. (Note that this
is the common reduced state of $\rho_{123}^f$ and
of the state (53).)

\section{Summing Up}

The investigation reported in this article is
restricted to von Neumann entropy, and von
Neumann mutual information defined by the
subsystem entropy decomposition $\enskip
S(\rho_{12})=
S(\rho_1)-I(\rho_{12})+S(\rho_2),\enskip$ where
$\rho_{12}$ is an arbitrary bipartite state, and
$\enskip \rho_i,\enskip i=1,2\enskip$ are its
reductions. The approach is based on the use of
coherence or incompatibility information $\enskip
I_C(A_2,\rho_i),\enskip i=2,12\enskip$ (cf
(1a)-(1c)), which quantifies the quantumness in
the relation of an observable and state.

Zurek's idea of "interrogating" the quantum
correlations of the composite state $\rho_{12}$
through the choice of a local observable $A_2$ is
elaborated via the mentioned subsystem
decomposition of entropy.

The first result (theorem 1 and (4a)) has
introduced coherence information into mutual
information through one of three relevant
non-negative terms. It is Zurek's discord
\cite{Zurek}, which turned out to be
coherence-information excess (global minus local)
(cf (7)). The other two terms are the information
gain and the residual mutual information.

The notion of function of observable or its
coarsening is made ample use of extending discord
also to incomplete observables. The second result
(theorem 2 and (9)) revealed that in refinement
(opposite of coarsening) both information gain
and discord are non-decreasing, and the residual
mutual information is non-increasing. It is known
from previous work \cite{Roleof} that coherence
information is non-decreasing in refinement. It
is somewhat surprising that also the (global
minus local) coherence-information excess (the
discord) is non-decreasing. (The finer observable
"sees" more quantumness both locally and in the
correlations; and the latter outweigh the
former.)

The zero-discord problem was explored in detail.
Two kinds of zero discord have been
distinguished: strong, when both terms in the
excess coherence information are zero, and weak,
when they are nonzero, but equal. Necessary and
sufficient conditions were given where possible.
Desirable results that have not been obtained
were pointed out.

A unique string of coarsenings of the
"interrogating" observable $A_2$ has been
derived: $\enskip D_2^{qc}\c C_2^{tw}\c B_2^{ess}
\c A_2,\enskip$ corresponding to (reading from
right to left) redundant noise, essential noise,
garbled information, pure quantum information and
pure quasi-classical information respectively
(see section IV.).

Finally, simplest possible measurement
interaction for measuring $A_2$ leading to a
tripartite state $\rho_{123}^f$, in which the
measuring apparatus is subsystem $3$, was
considered. The entropy relations in this state
were discussed. It was shown that all three terms
in the mutual information of $\rho_{12}$ are
shifted to the bipartite system $1+(2+3)$ in
$\rho_{123}^f$ (theorem 7 in section VI.).
Further, it was shown that the global coherence
information $I_C(A_2,\rho_{12})$ is shifted into
the global coherence information
$I_C(A_2,\rho_{123}^f)$ in $\rho_{123}^f$ (cf
(52)).

\begin{center}
{\bf Appendix A.}
\end{center}

Proof of theorem 2: In (4a) each $I(\rho_{12}^l)$
in the last term can be further decomposed
according to (4a) itself. Performing this and
substituting the result for each $l$ value in
(4a), one obtains the RHS of the claimed relation
(9).

To prove that the expression in the first large
brackets is the information gain, we write down
the decomposition of $S(\rho_1)$ due to probing
with $A'_2$ analogous to (6) in two versions:
directly and as a two-step procedure.
$$S(\rho_1)=\sum_{l,q}
\big[p_lp_{l,q}S(\rho_1^{l,q}||\rho_1)\big]+
\sum_{l,q}\big[p_lp_{l,q}S(\rho_1^{l,q})\big].
\eqno{(A.1)}$$ $$S(\rho_1)=\Big\{\sum_l\big[p_l
S(\rho_1^l||\rho_1)\big]+\sum_{l,q}\big[
p_lp_{l,q}S(\rho_1^{l,q}||\rho_1^l)\big]\Big\}+
\sum_{l,q}\big[p_lp_{l,q}S(\rho_1^{l,q})\big].
\eqno{(A.2)}$$ Comparison of (A.1) and (A.2)
proves the claim of theorem 2 as far as the
information gain with respect to $A'_2$ is
concerned.

It is obvious in (9) that the last expression is
the amount of inaccessible correlations. Since
the LHS is the same in (9) and (4a), the
expression in the second large brackets must be
the quantum discord. \hfill $\Box$

\begin{center}
{\bf Appendix B.}
\end{center}

We prove now the last claim in proposition 7. We
need auxiliary lemmata.

{\bf Lemma A.1:} If in a mixture of pure states
$\enskip \rho =\sum_{l=1}^mp_l\ket{l}\bra{l}
\enskip$ one has $p_{l=1}=r_{max}$, where
$r_{max}$ is the maximal eigenvalue of $\rho$,
then necessarily $\enskip \ket{l=1}\enskip$ is an
eigenvector of $\rho$ corresponding to the
eigenvalue $r_{max}$.

{\bf Proof:} It is known that for all $l$ values
$1=||p_l^{1/2}\rho^{-1/2} \ket{l}||^2,\enskip$
where $\enskip \rho^{-1/2}\enskip$ is the inverse
of the restriction of $\enskip \rho^{1/2}\enskip$
to the range of $\rho$ (\cite{Cassinelli}, see
Theorem 1 there). This implies $$ p_l=
\big(\bra{l}\rho^{-1}\ket{l}\big)^{-1}\quad
l=1,\dots ,m. \eqno{(A.3)}$$ (The operator
$\enskip \rho^{-1}\enskip$, by definition,
inverts the restriction of $\rho$ to its range.)

Let us expand $\enskip \ket{l}
=\sum_{k=1}^d\alpha_k^l\ket{r_k},\enskip
l=1,\dots ,m\enskip$ where $\enskip
\{r_k:k=1,\dots ,d\}\enskip$ is the positive
spectrum of $\rho$, and $\enskip
\{\ket{r_k}:k=1,\dots ,d\}\enskip$ is a
corresponding orthonormal set of eigenvectors.
Substituting this in (A.3), one obtains $$p_l=
\big(\sum_{k=1}^d|\alpha_k^l|^2r_k^{-1}\big)^{-1}\quad
l=1,\dots ,m. \eqno{(A.4)}$$ Assuming now that
$\enskip p_{l=1}=r_{max},\enskip$ one can write
$\enskip p_{l=1}^{-1}-r_{max}^{-1}=0, \enskip$
entailing with the use of (A.4)
$$\sum_{k=1}^d|\alpha_k^1|^2\big(
r_k^{-1}-r_{max}^{-1}\big)=0.$$ All terms are
nonnegative. This implies $\enskip
r_k<r_{max}\enskip \Rightarrow
\alpha_k^1=0.\enskip$ Hence, if $q$ enumerates
the possible multiplicity in $r_{max}$, then
$$\ket{l=1}=\sum_q\alpha_q^1
\ket{r_{max},q}.\eqno{(A.5)}$$ \hfill $\Box$

{\bf Lemma A.2:} If $\enskip \rho
=\sum_{l=1}^dr_l \ket{l}\bra{l}\enskip$  is a
mixture, and the weights $\enskip \{r_l:\forall
l\}\enskip$ coincide with the positive
eigenvalues of $\rho$ (with possible repetition
in the latter), then also the state vectors
coincide each with a corresponding eigenvector of
$\rho$: $\enskip \ket{l}=\ket{r_l},\enskip
l=1,\dots ,d.\enskip$ Naturally, $\enskip \rho
=\sum_lr_l\ket{r_l} \bra{r_l}\enskip$ is a
spectral form of $\rho$.

{\bf Proof:} We assume that in the mixture the
weights are written in non-increasing order.
Then, according to lemma A.1, $\enskip \rho
=r_{max}
\ket{r_{max}}\bra{r_{max}}+\sum_{l=2}^dr_l
\ket{l}\bra{l}.\enskip$ To apply total induction,
we further assume that the demonstration has
already been done up to $n$: $\enskip \rho =
\sum_{l=1}^nr_l\ket{r_l}\bra{r_l}+\sum_{l=n+1}^d
r_l\ket{l}\bra{l},\enskip$ where $\enskip 1\leq n
\leq (d-1).\enskip$ Let us introduce $\enskip
\beta_n\equiv \sum_{k=n+1}^dr_k,\enskip$ $\enskip
\beta_n>0$. Then $\enskip \rho'\equiv
\sum_{l=n+1}^dr_l/\beta_n \ket{l}\bra{l}=\rho
/\beta_n-\sum_{l=1}^nr_l/\beta_n\ket{r_l}\bra{r_l}
=\sum_{l=n+1}^d
r_l/\beta_n\ket{r_l}\bra{r_l}.\enskip$ The last
equality follows from the spectral form of
$\rho$. It is a spectral form of $\rho'$. Hence,
$\enskip r_{l=n+1}/\beta_n\enskip$ is its largest
eigenvalue. On account of lemma A.1, $\enskip
\ket{l=n+1}=\ket{r_{l=n+1}}.\enskip$ Total
induction then proves the claim of lemma
A.2.\hfill $\Box$

{\bf Lemma A.3:} If $\enskip \rho
=\sum_{l=1}^mp_l\ket{l}\bra{l}\enskip$ is a
mixture, and the so-called mixing entropy equals
the entropy of the state, i. e., $\enskip H(p_l)=
S(\rho ),\enskip$ then $\enskip m=d\enskip $, and
$\enskip \{p_l=r_l:l=1,\dots ,d\}\enskip$ is the
positive spectrum of $\rho$ (with possible
repetition in the eigenvalues).

{\bf Proof:} According to Theorem 3 in a
remarkable article by Nielsen \cite{Nielsen}, the
existence of the mixture in lemma A.3 implies
that its probability distribution is majorized by
the spectrum of $\rho$. This means that when both
$\enskip \{p_l:l=1,\dots ,m\}\enskip$ and
$\enskip \{r_k:k=1,\dots ,m\}\enskip$ are written
in non-increasing order (if $\enskip m>d,\enskip$
then $\enskip (m-d)\enskip$ zeros are added at
the end of the positive spectrum of $\rho$) then
$\enskip \sum_{l=1}^np_l\leq \sum_{l=1}
^nr_l,\enskip n=1,\dots ,(m-1).\enskip$

Next, we assume that the state space of $\rho$ is
at least $m$-dimensional. (If it is not, we can
orthogonally add a space to the null space of
$\rho$ without loosing generality of the
argument.) We define $\enskip \rho'\equiv
\sum_{k=1}^mp_l\ket{l}'\bra{l}',\enskip$ where
$\enskip \{\ket{l}':l=1,\dots ,m\}\enskip$ is an
arbitrary orthonormal set.

Ruch introduced the term "mixing character" for
the positive spectrum of $\rho$ (with possible
zeros) \cite{Ruch} (see also \cite{Lesche}), and
"larger" for the majorized spectrum. In a
previous article by the present author
\cite{FHII} the concept "strictly larger mixing
character" (when "larger" is not valid
symmetrically for the given mixing characters)
was treated, and it was shown that von Neumann
entropy is strictly mixing-homomorphic. This
means that if the mixing character of $\rho'$ is
strictly larger than that of $\rho$, then
$\enskip S(\rho')>S(\rho ).\enskip$

As it was stated, thanks to Nielsen, we know that
the mixing character of $\rho'$ is larger than
that of $\rho$. Since the entropies $\enskip
S(\rho')=H(p_l)\enskip$ and $S(\rho )$ are
assumed to be equal, the former cannot be
strictly larger. It must be equal. Mixing
characters are equal if and only if the
corresponding states have equal positive
eigenvalues with equal multiplicities. Hence,
$\enskip \{p_l=r_l:l=1,\dots ,m\}\enskip$ and
$m=d$ (the number of positive eigenvalues of
$\rho$ with possible repetitions) as claimed.
\hfill $\Box$

Finally, we prove the last part of proposition 7
claiming that if any mixture $\enskip
\rho=\sum_{l=1}^mp_l\rho_l\enskip$ of a finite
number of admixed states is given, and it has the
property that $\enskip S(\rho )=H(p_l)+
\sum_{l=1}^mp_lS(\rho_l),\enskip$ then the
mixture is orthogonal, i. e., $\enskip l\not= l',
p_l>0<p_{l'}\enskip \Rightarrow \enskip
\rho_l\rho_{l'}=0$.

Let $\enskip \rho =\sum_{l=1}^mp_l\rho_l\enskip$
be the given (initial) mixture. Let, further,
$\enskip \forall l:\enskip \rho_l=\sum_kr_k^l
\ket{lk}\bra{lk}\enskip$ be spectral forms.
Substitution in the initial mixture gives
$\enskip \rho
=\sum_l\sum_kp_lr_k^l\ket{lk}\bra{lk}\enskip$
with the mixing entropy $\enskip H(p_lr_k^l)=
H(p_l)+\sum_lp_lH(r_k^l)\enskip$ (cf (31b)).
Since by assumption $\enskip S(\rho )=H(p_l)+
\sum_lp_lS(\rho_l),\enskip$ and $\enskip \forall
l:\enskip S(\rho_l)=H(r_k^l),\enskip$ one has
$S(\rho )=H(p_lr_k^l).\enskip$ Hence, lemma A.3
is applicable to the mixture $\enskip \rho =
\sum_l\sum_kp_lr_k^l\ket{lk}\bra{lk},\enskip$
making $\enskip \{p_lr_k^l:\forall l,\forall k\}
\enskip$ the positive spectrum of $\rho$. Then
lemma A.2 implies that the mixture is a spectral
form of $\rho$. This cannot be unless the initial
mixture is orthogonal as claimed.\hfill $\Box$

\begin{center}
{\bf Appendix C}
\end{center}

Before we prove theorem 7, we establish some
facts. Since the operators at issue are twin
observables, one has
$$\forall l:\quad \rho_{123}^fP_2^l=\rho_{123}^f
\ket{l}_3\bra{l}_3\eqno{(A.6a)}$$ (cf the adjoint
of (34b) changing what has to be changed), and
equivalently (cf p. 052321-3 in \cite{FHPR02}),
$$\forall l:\quad P_2^l\rho_{123}^fP_2^l=
\ket{l}_3\bra{l}_3\rho_{123}^f\ket{l}_3
\bra{l}_3.\eqno{(A.6b)}$$ Besides, we need the
following result.

If $\rho_{12}$ is a bipartite density operator
and $\ket{b}_2$ is a second-subsystem unit
vector, then
$$\ket{b}_2\bra{b}_2\rho_{12}\ket{b}_2\bra{b}_2
=\Big[\tr_2\big(\rho_{12}\ket{b}_2\bra{b}_2\big)\Big]
\otimes \ket{b}_2\bra{b}_2\eqno{(A.7)}$$ (see the
necessity part in the proof of proposition 3).

{\bf Proof} of theorem 7:

{\bf a)} The validity of (48a) is a consequence
of (A.6a), (47b), and of the definition of
$p_l^f$ (see beneath (46b)).

{\bf b)} Utilizing (2c), (47c) and (A.6a), and
finally the definition of $\rho_1^{fl}$ (beneath
(46b)), and (48a), one has $$\forall l,\enskip
p_l>0:\quad \rho_1^l\equiv \tr_2(\rho_{12}^l)=
p_l^{-1}\tr_{23}(\rho_{123}^fP_2^l)=\rho_1^{fl}.
$$ This proves (48b).

{\bf c)} Claim (48c) is an immediate consequence
of definition (47a).

{\bf d)} Making use of (1a), of the mixing
property of entropy, and of (2b), one has
$$I_C(A_2,\rho_{12})=H(p_l)+\sum_lp_lS(P_2^l\rho_{12}
P_2^l/p_l)-S(\rho_{12})=$$ $$H(p_l)+\sum_lp_l
S(\rho_{12}^l)-S(\rho_{12}).\eqno{(A.8)}$$ On
account of (47c), (A.7) {changing what has to be
changed}, and (A.6b), one can write
$$S(\rho_{12}^l)=S\Big(\tr_3\big(\rho_{123}^f
\ket{l}_3\bra{l}_3\big)/p_l\Big)=
S\bigg(\Big[\tr_3\big(\rho_{123}^f
\ket{l}_3\bra{l}_3\big)/p_l\Big]\otimes
\ket{l}_3\bra{l}_3\bigg)=$$ $$ S\Big(\ket{l}_3
\bra{l}_3\rho_{123}^f\ket{l}_3\bra{l}_3/p_l\Big)
=S\Big(P_2^l\rho_{123}^fP_2^l/p_l\Big).$$
Substituting this in (A.8), making use of the
mixing property of entropy, taking into account
that
$$S(\rho_{12})=S(\rho_{12}\otimes
\ket{\phi }_3\bra{\phi }_3)= S\Big(U_{23}
(\rho_{12}\otimes \ket{\phi }_3\bra{\phi
}_3)U_{23}^{\dagger}\Big)= S(\rho_{123}^f),$$ and
utilizing (1a), one derives the RHS of (48d).

{\bf e)} To prove (48e), we argue in analogy with
the preceding item.
$$I_C(A_2,\rho_2)=H(p_l)+\sum_lp_l
S(P_2^l\rho_2P_2^l/p_l)-S(\rho_2).
\eqno{(A.9a)}$$ Further, (2b), (2c) and (47c),
upon taking partial trace $1$ of it, imply
$$S(P_2^l\rho_2P_2^l/p_l)=S\Big(p_l^{-1}\tr_3
(\rho_{23}^f \ket{l}_3\bra{l}_3)\Big) =
S\Big(p_l^{-1}\Big[\tr_3(\rho_{23}^f
\ket{l}_3\bra{l}_3\Big]\otimes
\ket{l}_3\bra{l}_3\Big).$$ Further evaluation
using (A.7) and (A.6b) after taking partial trace
$1$ in it gives
$$S(P_2^l\rho_2P_2^l/p_l)=S(P_2^l\rho_{23}^fP_2^l
/p_l).$$ Making use of the mixing property of
entropy once again, on account of $\tr
(\rho_{23}^fP_2^l)=p_l^f=p_l$ (cf beneath (46)
and (48a)), one obtains
$$S\Big(\sum_lP_2^l\rho_2P_2^l\Big)=
S\Big(\sum_lP_2^l\rho_{23}^f
P_2^l\Big).\eqno{(A.9b)}$$

Returning to the last term on the RHS of (A.9a),
one can write $$S(\rho_2)=
S\Big(U_{23}(\rho_2\otimes \ket{\phi}_3
\bra{\phi}_3)U_{23}^{\dagger}\Big)=S(\rho_{23}^f)$$
(cf (47a) upon taking $\tr_1$ in it).
Substituting this and (A.9b) in (A.9a), in view
of (1a), the RHS of (48e) is derived.

{\bf f)} Finally, to prove (48f), we write down
the definitions $$I(\rho_{12}^l)\equiv
S(\rho_1^l)+S(\rho_2^l)-S(\rho_{12}^l),
\eqno{(A.10a)}$$
$$I(\rho_{1,23}^{fl})\equiv
S(\rho_1^{fl})+S(\rho_{23}^{fl})-
S(\rho_{123}^{fl}).\eqno{(A.10b)}$$ The first
terms on the RHSs coincide due to (48b). Further,
utilizing $\tr_1$ of (47c),
$$S(\rho_2^l)=S\Big(\Big[\tr_3(\rho_{23}^f\ket{l}_3
\bra{l}_3)/p_l\Big]\otimes \ket{l}_3
\bra{l}_3\Big)=
S(P_2^l\rho_{23}^fP_2^l/p_l)=S(\rho_{23}^{fl})$$
(cf (A.7), (A.6b), and the definitions beneath
(46)). As to the third terms on the RHSs of
(A.10a) and (A.10b), equality is established by a
similar argument:
$$S(\rho_{12}^l)=S\Big(\Big[p_l^{-1}\tr_3(
\rho_{123}^f\ket{l}_3\bra{l}_3)\Big]\otimes
\ket{l}_3\bra{l}_3)\Big)$$ (cf (47c)). This
equals $S(\rho_{123}^{fl})$ (cf (A.7) and
(A.6b)).\hfill $\Box$

\end{document}